\documentclass[aps,prapplied,reprint,groupedaddress]{revtex4-2}
\usepackage{amsbsy}
\usepackage{amssymb}
\usepackage{amsmath}
\usepackage{graphicx}
\usepackage{dcolumn}
\usepackage{bm}
\usepackage{xcolor}

\begin{document}

\title{Very sensitive vapor-cell quasi-DC atomic E-field sensor}

\author{Amy Damitz, George Burns, Yuan-Yu Jau}
\affiliation{Sandia National Laboratories, Albuquerque, NM 87123, USA}

\date{\today}

\begin{abstract}
We report several technical approaches that significantly improve the performance of a vapor-cell atomic electrometer operating in the quasi-DC frequency domain ($\ll$ 1 kHz). With a very small active volume of approximately 11 mm$^3$ inside the vapor cell, we demonstrated a noise floor for electric field (E-field) sensitivity ranging from 0.2 to 7.7 mV/m$\sqrt{\rm Hz}$ for a frequency band of 1--100 Hz. Our work utilizes only a bare vapor cell for electrometry, without any metal parts or electrodes, to ensure minimal distortion of the measured E-field and to minimize the effective sensing volume for high spatial resolution. The E-field-sensitive atomic state (Rydberg state) is excited and read out optically, maximizing the simplicity of the system design and enabling the miniaturization of quasi-DC E-field sensors for potential applications, such as diagnostics of electronics without physical contact, communications in and below the super-low frequency (SLF) band, proximity detection, remote activity surveillance, tracing charge signatures, and research in bioscience and geoscience.
\end{abstract}


\maketitle

\section{Introduction}
Atomic Rydberg states with a large principal quantum number $n$ are electric-field (E-field) sensitive energy states that can interact with electric fields and cause noticeable energy level shifts due to their very-high polarizability that leads to strongly induced electric dipole moments. One of the major advantages of E-field sensing based on Rydberg atoms over other technologies is the establishment of traceable standards~\cite{Holloway2014IEEE}, since atoms are indistinguishable from one to another and the energy-level shift caused by the electric field is, in principle, determined only by fundamental constants. Over the past two decades, the research community working on Rydberg-enabled E-field sensing using atomic vapor cells has been focusing on the radio-frequency (RF) bands~\cite{Sedlacek2012,Holloway2014IEEE,Miller2016,Wade2018,cox2018,Paradis2019,Meyer2021,mingyong2020,Holloway2021,bang2022,rotunno2023,nowosielski2024,yang2024,wan2025,zhou2025,manchaiah2026,Zhang2026}. From a practical point of view, vapor-cell-based atomic devices illustrate many advantages over sensors that use ultra-high-vacuum (UHV) chambers for containing atoms. Vapor cells can hold large numbers of atoms in tiny, portable glass vessels. They do not require active vacuum pumps or other mechanisms to cool and trap the atoms/ions in a small volume, which largely mitigates the systematic complexity. In addition, vapor cells are easy to swap out with each other as replaceable parts without the need for substantial laser alignments.

While some advantages of of the calibration-free RF Rydberg E-field receivers are evident, the conventional electronic RF receivers still mostly lead in performance regarding sensitivity and detection bandwidth~\cite{ElectronicRFReceiver} owing to the fact that a high-gain, small-size antenna with low-noise RF electronics can be efficiently integrated for a frequency range from sub-GHz to tens of GHz. On the other hand, sensitively detecting electric fields in the quasi-electrostatic (quasi-DC) frequency domain ($\ll1$ kHz) with antenna dimensions much smaller than the wavelength becomes challenging. Thus, for quasi-DC E-field sensing applications that require good sensitivity and a very small sensing volume for high spatial resolution, atomic electrometry is more favorable than electronic RF receivers, since atoms are ``antennaless.''

The primary technical challenge of vapor-cell-based quasi-DC atomic electrometry is the low-frequency E-field screening effect on the vapor cell~\cite{Jau2020}, which was first reported by Mohapatra \emph{et al.}~\cite{mohapatra2007}. Due to the laser availability and the desired vapor pressure, alkali-metal atoms remain the main atomic species used in vapor-cell Rydberg electrometers. These metallic atoms eventually form a thin ``film'' on the inside of vapor cells. The thin layer of metallic atoms provides some minimal electrical conductivity and acts as an imperfect Faraday cage that allows high-frequency electric fields to pass through while effectively blocking external electric fields of sufficiently low frequencies. This is an important reason why the mainstream Rydberg E-field sensing community concentrates more on the RF bands. In 2019, our team at Sandia National Laboratories (SNL) made an important breakthrough by greatly reducing the E-field screening problem using a sapphire-walled rubidium (Rb) vapor cell to demonstrate atomic electrometry for E-field detection at frequencies below 1 kHz, with sufficiently good sensitivity to detect electric fields emanating from moving charged objects and permeating the vapor cell to directly interact with the Rydberg atoms~\cite{Jau2020}. There have been other efforts to explore the regime of detecting sub-kHz electric field outside of the alkali-metal-atom container by using a pair of electrodes to connect the internal space containing the atomic vapor with the external space out of the glass cell~\cite{holloway2022,li2023,lei24,arumugam2025,han2025,xiao2025}. Implementing electrodes inside the vapor cells, however, does not resolve the E-field screening problem. The finite electrical conductivity on the inner surface of the glass cell still produces non-negligible electrical connection between the electrodes, which impedes low-frequency detection due to the large source impedance $Z_s=1/i\omega C_s$ especially when the detection angular frequency $\omega$ is low with a small source capacitance $C_s$ of the electrode connectors outside the vapor cell. The high source impedance hinders the efficiency of coupling the electric fields from free space to the internal electrodes, unless a larger external antenna is assigned to the electrode connectors to effectively reduce the $Z_s$. The effective sensing volume is then determined by the antenna dimensions, which are, inevitably, increased. Therefore, most devices using electrodes inside vapor cells can be considered as signal voltage meters, since relatively low-impedance voltage or signal sources are used to drive the internal electrodes to generate more well-defined electric fields for interacting with Rydberg atoms, rather than using the electric field directly originating from external space.

To continuously improve the sensitivity of atomic E-field sensors using bare vapor cells in the quasi-DC regime, we must further tackle the E-field screening issue in alkali-vapor cells. In this paper, we report our experimental studies on materials and coatings for manufacturing suitable vapor cells for E-field sensing. We present several technical approaches and experimental results to largely enhance the sensitivity of the vapor-cell atomic electrometer over a frequency range from 1 to 100 Hz, achieving more than an order of magnitude improvement compared to our previous work~\cite{Jau2020}. This newly demonstrated quasi-DC electrometry therefore surpasses electronic E-field detection technology when considering a similarly small effective sensing volume within the same frequency range.

\begin{figure}[t]
 \begin{centering}
 	  \includegraphics[width=0.45\textwidth]{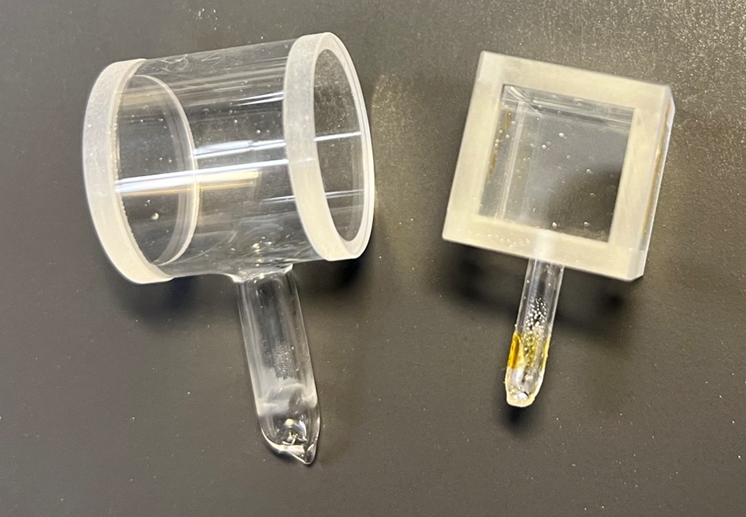}\\

  \caption{The two major geometries of the vapor cells used in our experimental work are shown here. On the left is the cylindrical Rb vapor cell made with quartz or Pyrex glass, with and without coatings, and on the right is the cubic Rb vapor cell made with monocrystalline sapphire. The cylindrical cells have an external diameter of 25 mm and a length of 25 mm, and the window thickness is 3 mm. The cubic cells have external dimensions of $20\times20\times20$ mm$^3$ with a wall thickness of 3 mm.}
        \label{fig:vaporcell}
 \end{centering}
\end{figure}
\section{Cell materials and cell coatings}
One of the main roadblocks in using alkali-vapor cells for quasi-DC E-field sensing is the deposition and interaction of alkali metals on the inner surfaces of the vapor cells, which interfere with the ability of Hz and kHz electric fields to penetrate through the glass/alkali-metal surface with non-zero electrical conductivity. While the merit of a monocrystalline sapphire-made alkali-vapor cell has been identified as being able to greatly mitigate the E-field screening issue~\cite{Jau2020}, the cost and intricacy of manufacturing a sapphire-made vapor cell hampers broader development of atomic E-field sensors. Thus, we took an empirical approach in searching for alternatives for making vapor cells that can have a lower cost and might deliver relevant characteristics that are close to or better than sapphire-made vapor cells.

In this experimental study, we performed measurements of the E-field screening rate, $\gamma$, using the two-photon technique as described in Section V of Ref.~\cite{Jau2020} to determine the effective sheet resistance $R_\Box$ on the inner surfaces of different vapor-cell samples, which is independent of the geometry and material of the cell. Here, we assume $R_\Box$ is uniformly distributed inside a vapor cell for the definition of ``effective.'' In addition to the sapphire-made cells, we prepared glass cells made with quartz or Pyrex glass, which are more conveniently manufactured through conventional glassblowing techniques, and applied different coatings inside the cells. Figure~\ref{fig:vaporcell} illustrates two cell geometries used in this work. We selected coating materials including amorphous Al$_2$O$_3$ (alumina), diamond-like coating (DLC), MgO, and MgAl$_2$O$_4$. These coatings were chosen based on alkali resistance, according to a thesis by Fletcher~\cite{flectcher2017}, along with which coatings could be obtained and applied to the inner surface of a vapor cell via atomic layer deposition (ALD) or plasma-enhanced chemical vapor deposition (PECVD). Inspired by our previous work~\cite{Jau2020}, we also tested anti-spin-relaxation coating materials, paraffin and octadecyltrichlorosilane (OTS), which can be directly ordered from the glass cell vendor.

\begin{figure}[t]
 \begin{centering}
 	  \includegraphics[width=0.45\textwidth]{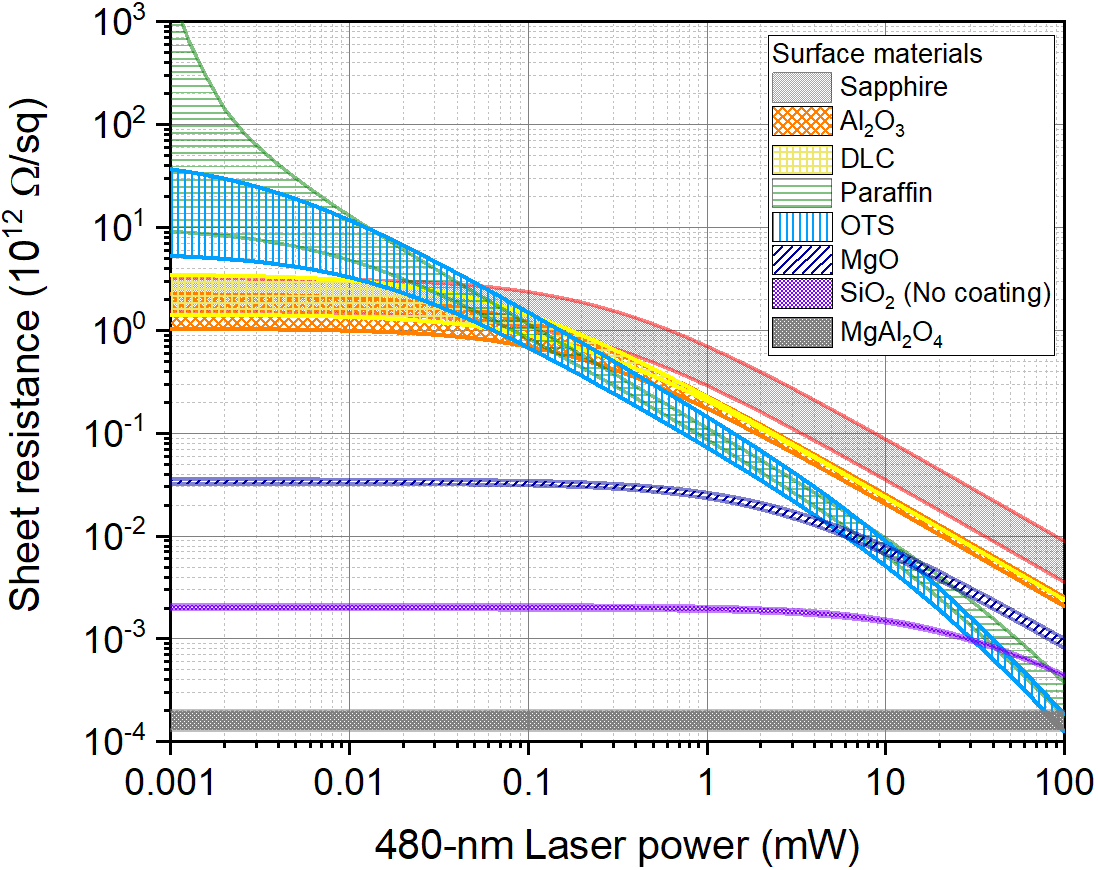}\\

  \caption{Sheet resistance $R_\Box$ for different surface materials over 480-nm laser power with these measurement conditions: diameter of the laser beams =1.6 mm; 780-nm laser power = 25 $\mu$W; Rb number density $\approx1.4\times10^{10}$ cm$^{-3}$ (measured); cell-body temperature $\sim45$ $^\circ$C; bias magnetic field $\sim7$ G.}
        \label{fig:2photoncoat}
 \end{centering}
\end{figure}

From our past research, we are aware that $R_\Box$, or surface free charges for electrical conduction, can be affected by several control parameters. As a rule of thumb, increasing laser power, increasing Rb vapor pressure, or reducing the inner surface temperature always leads to a decrease in $R_\Box$ and vice versa. We maintained approximately the same measurement conditions for all cell samples, and the measurement results of $R_\Box$ for different surface materials over 480-nm laser power are summarized in Fig.~\ref{fig:2photoncoat}. Instead of using a mathematical approximation of $R_\Box\approx(\gamma\epsilon V^{1/3})^{-1}$~\cite{Jau2020}, where $\epsilon$ is the effective electric permittivity, and $V$ is the volume of the vapor cell, we used time-dependent finite-element modeling (FEM) and experimental measurement of $\gamma$ values~\cite{Jau2021r} to derive $R_\Box$ with better accuracy. In Fig.~\ref{fig:2photoncoat}, the area of each curved band represents the experimental uncertainty and the variation for vapor cells with the same inner-surface materials. The original $\gamma$ data for each surface material over 480-nm laser power was fitted linearly with a zero-power offset, except for the paraffin and OTS data that required quadratic curve fitting. At zero 480-nm laser power, paraffin and OTS deliver the highest sheet resistance as conjectured in Ref.~\cite{Jau2020}. With higher 480-nm laser power, $R_\Box$ falls more rapidly with paraffin and OTS compared to others, which may be associated with light-induced atomic desorption~\cite{Karaulanov2009}, and sapphire shows the slowest reduction in $R_\Box$ as the 480-nm power increases. In practice, we must apply a sufficient amount of Rydberg excitation laser power (the 480-nm power in this case), and therefore, we are more interested in the results with 480 nm power above 10 mW. We find that sapphire is still the best, and Al$_2$O$_3$ and DLC are only about a factor of 2--4 worse than sapphire, which is still much better than the bare SiO$_2$ glass surface. Hence, Al$_2$O$_3$ and DLC are suitable coating options for making cost-effective vapor cells for quasi-DC E-field sensing. Whether the cell was made of Pyrex glass or quartz did not affect the results with the DLC or Al$_2$O$_3$ coating. The only coating that was worse than no coating at all was MgAl$_2$O$_4$, for which accurate E-field screening rates were too fast to be measured, limited by the shortest time constant of the lock-in amplifier used in the experiment. The outcome of the MgAl$_2$O$_4$ coating may not be valid. It is possible that something went wrong during the coating process. In summary, although the characterization results for different surface materials were acquired with only one experimental circumstance, the relative ranking of the $R_\Box$ for each material should remain the same with other testing conditions. Recently, the theoretical work by Ma \emph{et al.} indicates that CaF$_2$~\cite{ma2025} may also be a good material for low E-field screening in the vapor cell.

\section{Technical schemes for vapor-cell quasi-DC atomic electrometry}
Since the first demonstration of atomic electrometry below 1 kHz with a bare vapor cell~\cite{Jau2020} at SNL, we have discovered and conceived new technical approaches to further improve E-field sensitivity across the frequency bands from super-low frequency (SLF) and extremely low frequency (ELF) to nearly DC. Here, we discuss four primary technical schemes~\cite{Jau2022r}: using a strong bias magnetic field (B-field) to suppress the E-field screening effect, using three near-infrared laser fields to interrogate a Rydberg resonance, using a Rydberg $P$-orbital for electric field sensing, and using an externally switched electric field to eliminate the need for LED-induced E-field biasing inside the vapor cell.

\subsection{Magnetic-field suppressed E-field screening}
\begin{figure}[b]
 \begin{centering}
 	  \includegraphics[width=0.45\textwidth]{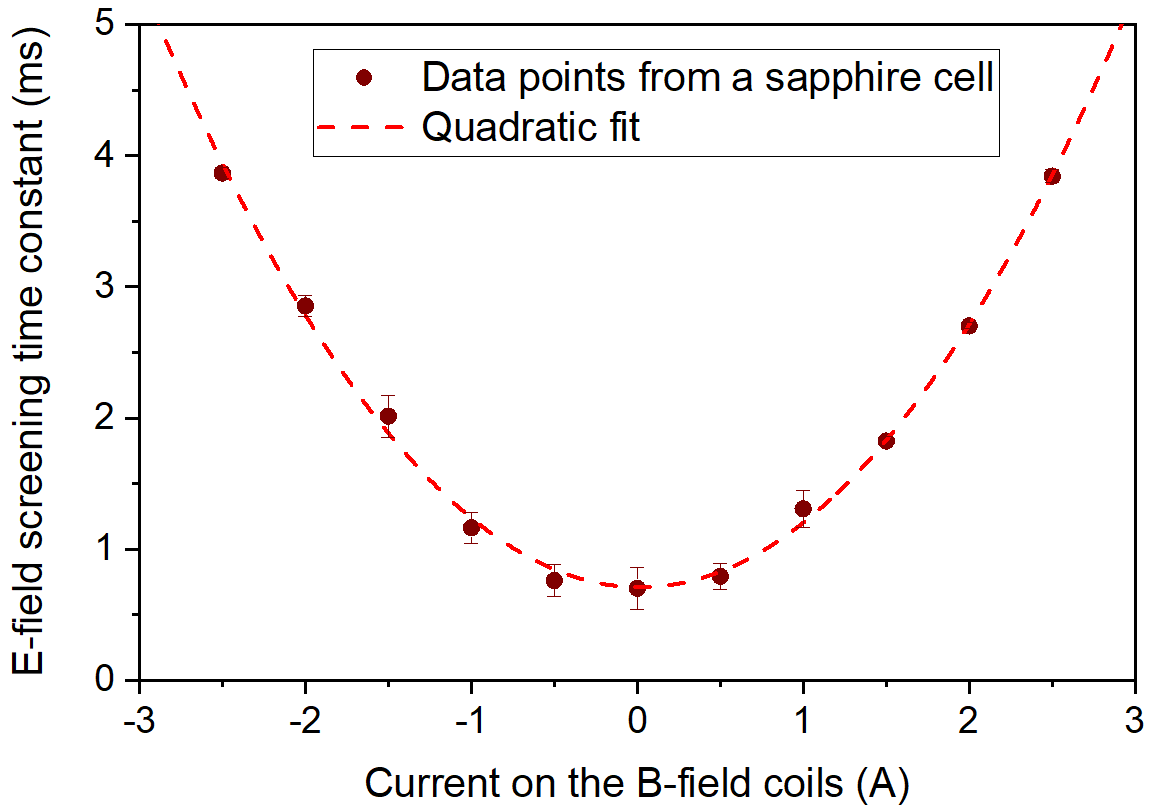}\\

  \caption{Example data of the experimentally verified B-field suppressed E-field screening. The screening time constant $\tau$ quadratically depends on the B-field coil current. The B-field amplitude vs. the coil current ratio is 9.7 G/A. The corresponding experimental conditions are: laser beam diameter = 1 mm; 480-nm laser power = 100 mW; 780-nm laser power = 25 $\mu$W; Rb number density $\approx4\times10^{10}$ cm$^{-3}$; cell-body temperature $\sim35$ $^\circ$C.}
        \label{fig:2photonbfiel}
 \end{centering}
\end{figure}
For vapor-cell-based atomic electrometry, we typically apply a bias B-field to define the quantization axis for laser interrogations. The B-field strength is usually chosen to be a few gauss (G). Throughout a series of experiments measuring the E-field screening rates, we surprisingly discovered that the E-field screening time constant $\tau=1/\gamma$ could be quite sensitive to the driving current on the B-field coils~\cite{Jau2021r}. After more systematic studies, we found that $\tau$ demonstrates a quadratic dependence on the B-field amplitude within a certain range of the B-field strength, as illustrated in Fig.~\ref{fig:2photonbfiel}. This phenomenon can be observed in sapphire cells and regular glass cells with paraffin, OTS, Al$_2$O$_3$, DLC, and MgO coatings across a broad range of applied laser power, with the only change in the B-field dependence being the quadratic coefficient for different experimental settings.

This ``magnetoresistance-like'' effect was not publicly reported, and its exact physical mechanism has not yet been identified. The B-field may influence the sheet resistance of the inner surface, the surface photoelectric effect, the transport of charged particles, such as positive atomic ions and negative electrons, and the ionization processes. The B-field may have interplay in one or more of these possible scenarios. Although we are not certain about the underlying physics for this new discovery, since the E-field screening rate is inversely proportional to the square of the bias B-field amplitude, increasing the B-field strength can allow better penetration of the low-frequency electric fields into the vapor cell, providing a significant advantage in exploiting this unexpected B-field effect to enhance quasi-DC atomic E-field sensing if the signal-to-noise ratio (SNR) remains unchanged. We have verified that the B-field suppressed E-field screening is still present with minimal Rydberg excitation using very low laser power. Whether this magnetoresistance-like phenomenon is optically associated or not, it represents a strong effect. Most previously discovered magnetoresistance mechanisms require a B-field strength of at least a few hundred G to several tens of thousand G, and some cases require cryogenic temperature to show changes in electrical resistance from a few percent to several hundred percent~\cite{ramirez1997,ennen2016,niu2022,ritzinger2023}. In our case, the B-field strength is in a much lower range (a few to several tens of G) at room temperature, and we observe a greater than 100\% change in surface resistance. This effect may lead to other useful applications~\cite{Jau2022r}.

\subsection{Three-photon Rydberg interrogation}
\begin{figure}[t]
 \begin{centering}
 	  \includegraphics[width=0.48\textwidth]{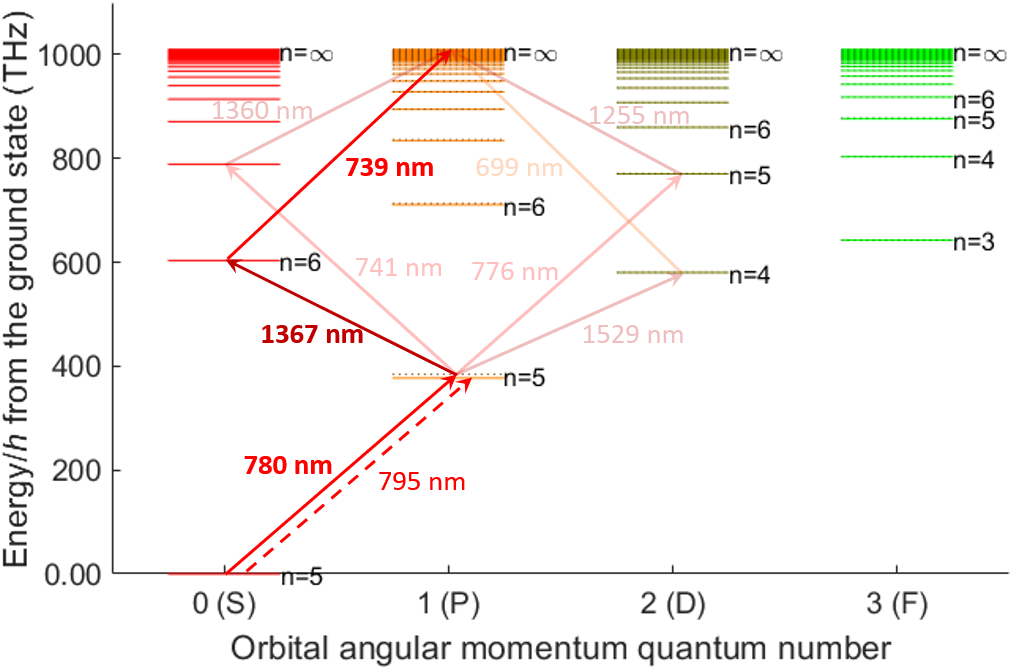}\\

  \caption{Rb energy-level diagram with some possible three-photon excitation methods to reach Rydberg $P$ states. The highlighted wavelengths are used in this work.}
        \label{fig:3photonenergylevel}
 \end{centering}
\end{figure}

Two-photon electromagnetically induced transparency (EIT) spectroscopy for quasi-DC electrometry suffers from a rapidly increasing E-field screening rate (corresponding to a decreasing sheet resistance) that is linearly proportional to the 480-nm laser power, as indicated in Fig.~\ref{fig:2photoncoat}. Therefore, while boosting the 480-nm laser power can increase the SNR, it also raises the low-cutoff frequency $f_{\rm 3dB}$ of the high-pass E-field screening effect in the vapor cell. Hence, only the high-frequency sensitivity is improved, while the low-frequency sensitivity remains the same but its dynamic range of signal detection does increase. The 480-nm laser obviously constrains the quasi-DC E-field sensing performance. In the presence of laser beams, we believe that additional free charges on the inner surface of a vapor cell come from two possible sources. One source is the generation of photoelectrons due to the photoelectric effect on the inner surface, and the other is the ionization of Rydberg atoms, which can supply charged particles to the cell's inner surface. Considering the possible work function of Rb atoms on the surface, the 480-nm laser can easily excite photoelectrons from the inner surface, while the 780-nm laser cannot~\cite{WorkFunctionNote}. These photoelectrons may travel from one surface area to another surface area and/or interact with Rb atoms before returning to the inner surface. The Rb Rydberg atoms can further be photoionized or electroionized by lasers or free-space electrons. The ionization-produced Rb$+$ and e$-$ can move freely and be transported to the inner surface. In addition, collisions of Rydberg atoms with the cell's inner surface may also generate surface free charges. The exact contributions from all of the possible mechanisms to the laser-enhanced E-field screening in the vapor cell are undetermined, but it is certain that the optically assisted generation of free charges due to energetic photons from the short-wavelength laser plays an important role. Thus, if we can avoid using short-wavelength lasers, such as the 480-nm laser, this should significantly minimize the optically promoted E-field screening effect. Therefore, we propose to use a three-photon scheme with all near-IR lasers~\cite{Jau2022r} to interrogate Rydberg $P$ orbitals, as depicted in Fig.~\ref{fig:3photonenergylevel}, to further reduce the E-field screening in the vapor cell in the presence of lasers. Our approach has a different motivation from other efforts in the literature~\cite{Ryabtsev2011,Ripka2021} of using three-photon Rydberg interrogation for achieving different advantages in RF electrometry.

\begin{figure*}[t]
 \begin{centering}
 	  \includegraphics[width=\textwidth]{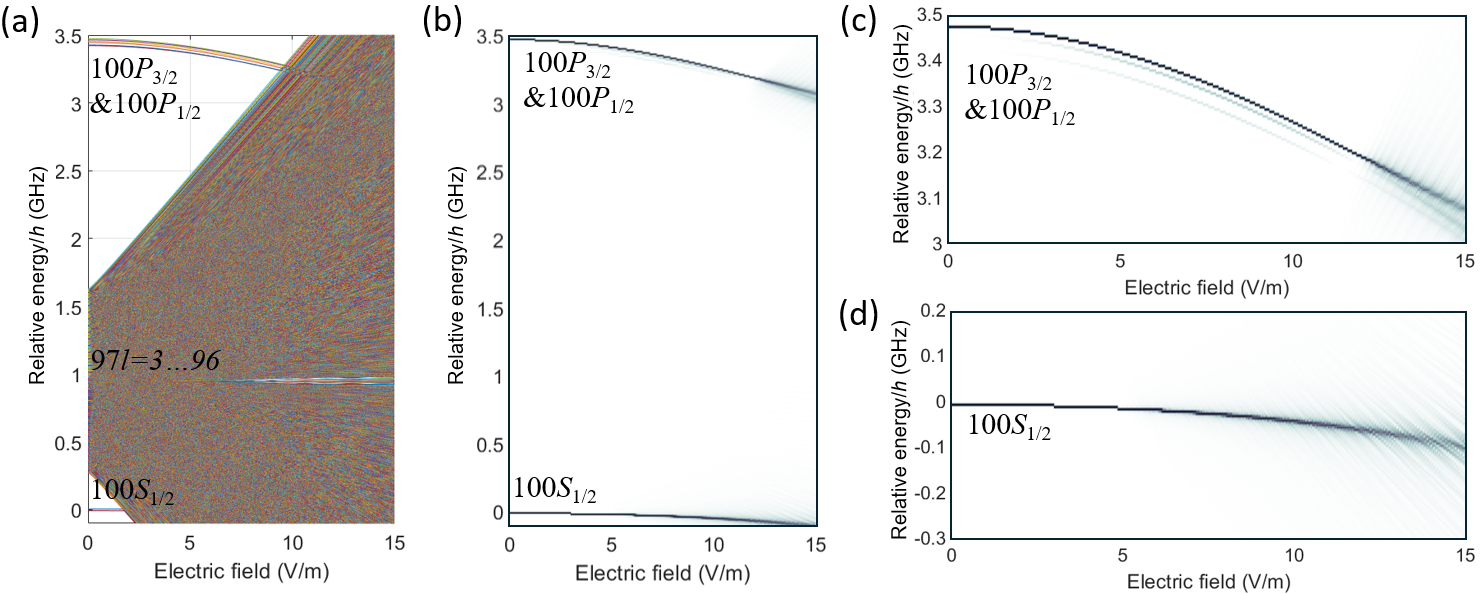}\\

  \caption{(a) Energy levels as functions of electric field with a bias B-field at  5 G for $n=100$, $l=0,1$ and $n=97$ and $l=3\ldots96$. (b) Relative coupling strengths (expressed in darkness) of the $100S$ and $100P$ orbitals to the interrogation lasers. (c) Zoom-in of the $100P_{3/2}$ and $100P_{1/2}$ E-field induced shift states. (d) Zoom-in of the $100S_{1/2}$ E-field induced shift states. One can clearly see that the $100S_{1/2}$ state is less sensitive to the electric field and experiences interference from high-order orbitals at a lower electric-field strength compared to the $100P_{3/2}$ and $100P_{1/2}$ states.}
        \label{fig:100pvss}
 \end{centering}
\end{figure*}
In a three-photon scenario, a Rydberg $P$ state can be reached via two intermediate states. One suitable combination of laser wavelengths used in our work is with 780 nm, 1367 nm, and 739 nm, which are all in the near-IR range and do not have enough photon energy to directly engender additional surface free charges and photoelectrons. The excitation path through atomic electronic states is from the upper hyperfine manifold of the 5$S_{1/2}$ ground state via the highest hyperfine manifold in the 5$P_{3/2}$ state, then to the upper hyperfine manifold of the 6$S_{1/2}$ state, and finally reaching the $nP_{3/2}$ Rydberg state, as shown in Fig.~\ref{fig:3photonenergylevel}. Owing to the use of the 6$S_{1/2}$ intermediate state, a repump laser at 795 nm~\cite{Jau2021d} must be applied to prevent the Rb atoms from optically pumping into the dark state, the lower manifold of the ground-state hyperfine structure. Although the three-photon scheme seems to complicate the laser configuration, in practice, it is quite manageable. In developing a practical quasi-DC atomic E-field sensor system, the benefits of the three-photon scheme outweigh the minor burden of the additional laser compared to the two-photon scheme. We identify the primary advantages of the three-photon Rydberg interrogation as follows: (1) a significant reduction in the E-field screening rate for E-field frequencies between DC and the SLF band from using just near-IR range lasers and eliminating the short-wavelength laser, (2) advantageous Rydberg $P$ orbitals (with more details in Section C), and (3) the utilization of mature semiconductor near-IR lasers.

\subsection{Rydberg $P$ orbital}
Both Rydberg $S$ and $D$ orbitals can be probed via two-photon EIT Rydberg spectroscopy. Rydberg $P$ and $F$ orbitals can be reached via a three-photon approach. We carried out some detailed calculations with similar numerical modeling methods described in Ref.~\cite{Happer2010} and used in Refs.~\cite{Jau2016,Jau2020} to understand the E-field influence on these Rydberg states, and the results are plotted in Fig.~\ref{fig:100pvss}. The 100$S_{1/2}$ state and the 100$P_{3/2}$ state are of particular interest because the former was used in previous work~\cite{Jau2020}, while the latter is what we propose to use. One can see that the energy of the Rydberg $S$ orbital is closer to that of the high-order orbitals with angular momentum quantum number $l\geq3$, where $l=0,1,2,3,\cdots$ for $S$, $P$, $D$, $F$, $\cdots$ orbitals. Therefore, a high-$n$ Rydberg $S$ state experiences interference from the high-order orbitals quickly when raising the electric-field amplitude as illustrated in Fig.~\ref{fig:100pvss}. For the 100$S_{1/2}$ state in the ``spaghetti'' region (made of $n=97$ and $l=3\cdots96$ quantum states), the atom not only delivers an effectively broader resonance linewidth due to level mixing but can also evolve into high-order angular momentum states if the electric-field condition changes, and thereafter the atom can no longer contribute to the signal. This limits the maximum applicable bias E-field and restricts the linear E-field shift coefficient $\kappa$ (i.e., the slope of the Stark-shift curve), which is proportional to the amplitude of the bias E-field (with more details in Section D). One can see that compared to the 100$S_{1/2}$ state, the 100$P_{3/2}$ state not only can extend further before hitting the spaghetti levels, but its polarizability is also about six times stronger, and therefore the E-field sensitivity can be, in principle, six times better with the same E-field biasing strength.

\subsection{Internal biasing using external switching E-field}
\begin{figure}[t]
 \begin{centering}
 	  \includegraphics[width=0.48\textwidth]{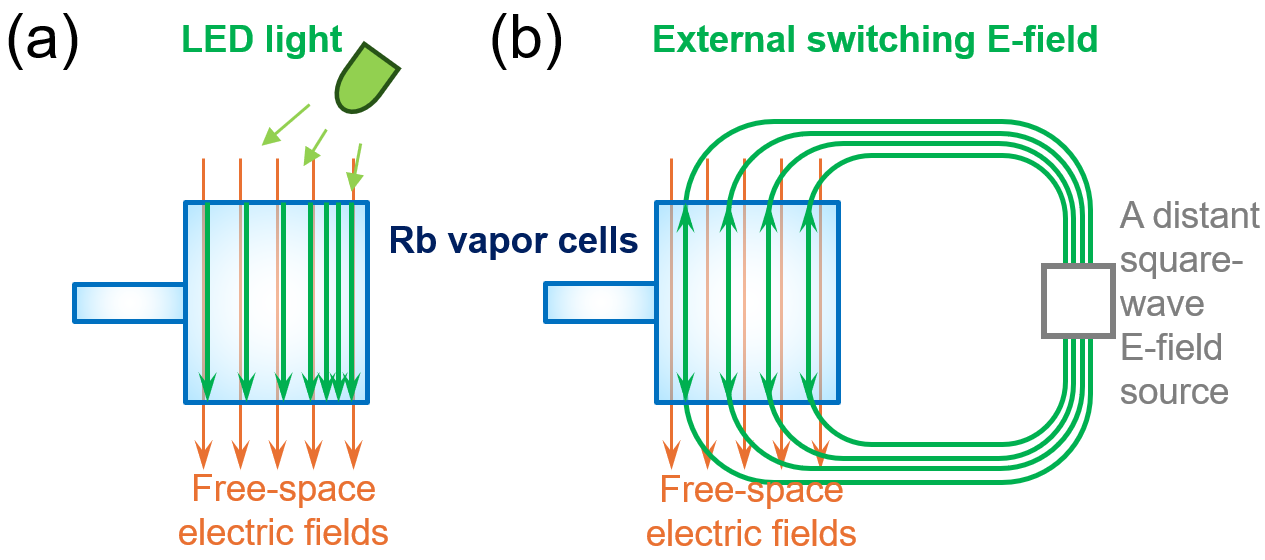}\\

  \caption{(a) The previous approach of generating static internal bias E-field inside a vapor cell using LED light. (b) The new approach of using an external switching bias E-field that can lead to a more uniform E-field biasing inside the vapor cell.}
        \label{fig:E-fieldBiasing}
 \end{centering}
\end{figure}

Rydberg associated electric dipole moments of a selected Rydberg state come from state mixing with other Rydberg states, indexed by $i$, that have opposite parity. The associated electric dipole moment from each pairwise state mixing has a characteristic frequency $\Delta_i/2\pi$, where $\hbar\Delta_i$ is the energy difference of the pair of states. We find the E-field induced frequency shift of the selected Rydberg state to be $\nu_s=(2\pi)^{-1}\sum_i(\mathbf{D}_i\cdot\mathbf{E})^2/(4\Delta_i\hbar^2)\approx\alpha E^2$ for $|\mathbf{D}_i\cdot\mathbf{E}|<|\Delta_i\hbar|$, the so-called Stark shift, where $\mathbf{D}_i$ is the vector dipole moment via mixing a selected pair of Rydberg states, $\mathbf{E}$ is the vector electric field, and $\alpha\propto n^7$ is the quadratic E-field shift coefficient, which is essentially the polarizability with units of Hz/(V/m)$^2$. This quadratic dependence, however, makes the Rydberg resonance frequency $\nu_s$ insensitive to a very small change $\delta E$ in a quasi-DC E-field because $\delta\nu=\alpha \delta E^2$. In order to increase the sensitivity of $\delta E$ and achieve a linearly dependent frequency shift $\delta\nu_s=\kappa\delta E$, we can introduce a constant bias E-field, $E_b$. Therefore, the total E-field is $E=E_b+\delta E$, and we find $\delta\nu_s=2\alpha E_b\delta E$ for $\delta E\ll E_b$, and thus, $\kappa=2\alpha E_b$. The sensitivity is then enhanced by a factor of $2E_b/\delta E$ compared to the quadratic E-field response. This is a heterodyne concept at DC as introduced in our previous work~\cite{Jau2020}. Owing to the low-frequency E-field screening effect, an external static bias E-field cannot enter a vapor cell. To mitigate this challenge, we have been using the LED light-induced charge patches on the vapor-cell inner surface to generate a bias E-field as shown in Fig.~\ref{fig:E-fieldBiasing}a. The main practical issue of the LED induced bias E-field is the spatial inhomogeneity of $E_b$, which denotes as $\Delta E_b$. Hence, the effective Rydberg resonance linewidth is $\Delta\nu_{\rm eff}=\Delta\nu\oplus2\alpha E_b\Delta E_b$, where $\Delta\nu$ is the intrinsic linewidth, and $\oplus$ denotes a convolution. Thus, the sensitivity level $(\Delta\nu_{\rm eff}/\kappa)/{\rm SNR}$ is $(\Delta\nu⁄2\alpha E_b)/{\rm SNR}$ for a small $E_b$ and $\Delta E_b/{\rm SNR}$ for a large $E_b$. One can see that increasing the bias E-field stops improving the sensitivity when $2\alpha E_b\Delta E_b$ becomes much larger than $\Delta\nu$.

From a past experiment, we learned that LED induced bias E-field has inhomogeneity $\Delta E_b/E_b\sim100$\%, a quite unfavorable situation. After all, we do not have a definitive way to exactly control the distribution of charge patches on the inner surface of a vapor cell created by the LED light. To mitigate this problem, we can use a switching E-field to generate a bias E-field inside the vapor cell with an assigned direction. In this approach, a more uniform E-field can be produced from an external source at a distance from the cell as illustrated in Fig.~\ref{fig:E-fieldBiasing}b. This idea was experimentally validated with a two-photon scheme~\cite{Jau2024r}. The field polarity alternates at a rate faster than the E-field screening rate, and naturally this external switching E-field can enter the vapor cell. Using FEM simulations for ideal scenarios, we find the bias E-field inhomogeneity along the laser beam path to be about 11\% and 7\% for sapphire and quartz cells, respectively, which is far more uniform than the LED bias E-field. This method enables the vapor cell to utilize a higher $E_b$ to attain a larger $\kappa$. With this scheme, when the E-field variation $\delta E=0$, either $+E_b$ or $-E_b$ causes the same frequency shift $\nu_s$. When $\delta E\neq0$, we then see an oscillation in $\nu_s$ at the switching frequency with an amplitude $\propto2\alpha E_b\delta E_b$, which can be demodulated using a second lock-in amplifier, while the first lock-in is used for demodulating the dithered Rydberg excitation laser for generating a dispersive-like Rydberg resonance signal~\cite{Jau2020}. Additionally, a well defined orientation of $E_b$ can enable a reconstruction of the 3D-vector information of the electric field by measuring E-field amplitude along three orthogonal directions.

\begin{figure*}[t]
\begin{centering}
 	  \includegraphics[width=1.08\textwidth]{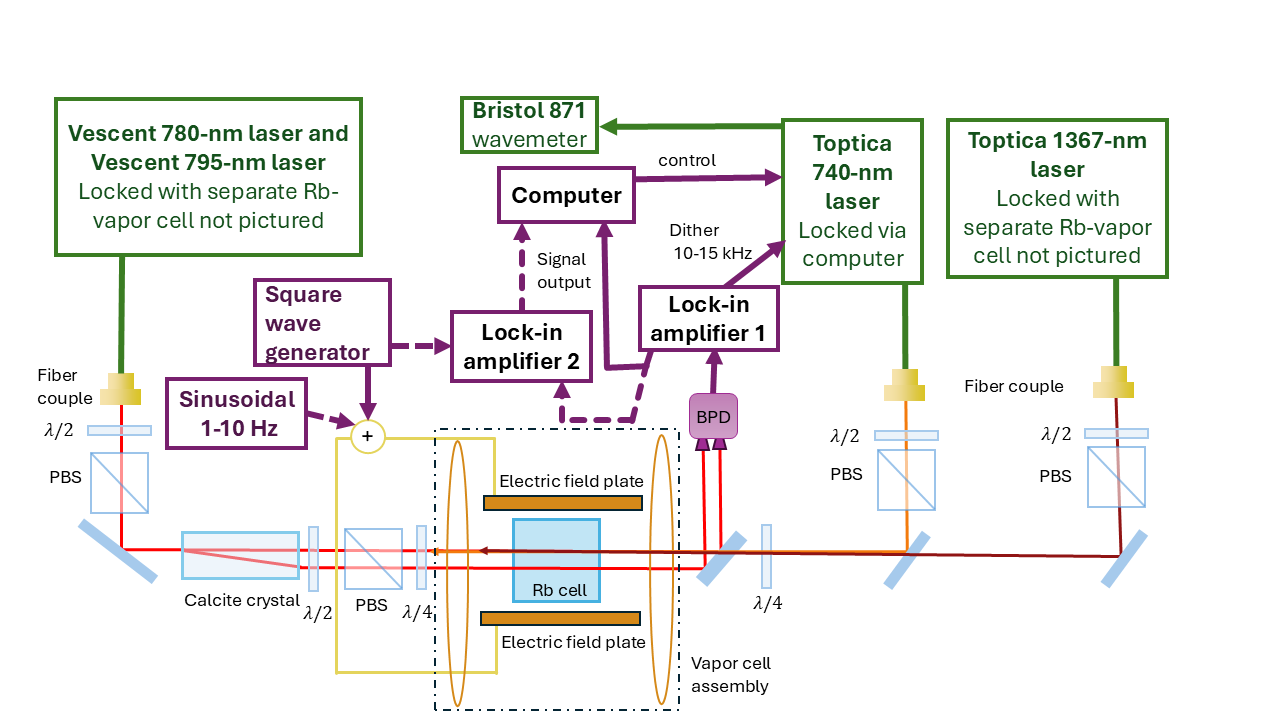}\\

  \caption{Experimental setup for observing the three-photon Rydberg resonance, characterizing E-field screening on the vapor cell, and determining quasi-DC E-field sensitivity. PBS, polarizing beam splitter cube; $\lambda/2$, half-wave plate; $\lambda/4$, quarter-wave plate; BPD, balanced photo detector.}
        \label{fig:setup}
\end{centering}
\end{figure*}
\section{Experimental implementation}
To validate the technical approaches described in the previous sections, we establish a tabletop setup for the three-photon experiment as illustrated in Fig.~\ref{fig:setup}, which is quite similar to the two-photon setup described in our previous work~\cite{Jau2020}. Three lasers at 780 nm, 1367 nm, and 739--741 nm are used to excite the Rb atoms into an $nP_{3/2}$ Rydberg state, while the 795 nm laser is used as a repump, as described in Section B. For the experiments presented in this paper, we use the $^{85}$Rb isotope.

All of the lasers are coupled into polarization-maintaining (PM) fibers to be delivered to the vapor cell. The 780 nm and 795 nm lasers are both coupled into the same fiber, while the 1367 nm and 740 nm lasers each have their own separate fibers. The 780/795 nm laser is split into two parallel, equal-power beams by a calcite crystal and a half-wave plate (HWP). Both of the 780/795 nm laser beams pass through the Rb vapor cell. The 1367 nm and 740 nm laser beams are overlapped using a dichroic mirror and then counter-propagated along one of the 780/795 nm beams. The sizes of all four laser beams vary individually from approximately 0.5 mm to 2 mm in diameter, depending on experimental tasks and specific optimizations. The beam diameters were mainly chosen to increase the SNR of the Rydberg resonance. All of the lasers have a HWP, a polarizing beam-splitting (PBS) cube, and a quarter-wave plate (QWP) to circularly polarize them before entering the vapor cell. The circular polarization direction for the 780/795 nm laser, 740 nm laser, and 1367 nm laser is chosen such that the atoms are excited into the correct state in the transition ladder, as seen in Fig.~\ref{fig:3photonenergylevel}. To allow only the 780 nm light into the balanced photodetector (BPD) while counter-propagating the 740 nm and 1367 nm lasers, a filter is used that transmits the 740 nm and 1367 nm light while reflecting the 780/795 nm light. On the BPD, there are band-pass filters that allow only 780 nm beams to pass through to the two photodiodes on the BPD. To maintain the long-term stability of the lasers, the 780 nm, 795 nm, and 1367 nm lasers are all frequency-locked to the desired atomic transitions using reference vapor cells. We sweep the optical frequency of the 740-nm laser near the value calculated by the quantum defect method~\cite{Jau2020}, which can be verified by a wavemeter, to find the resonance signal of an $nP_{3/2}$ state. Over the course of the experiment, we use different $n$ levels depending on the requirements of the experiment. For instance, the main $nP_{3/2}$ states used are the $n=50$, which is used when characterizing E-field screening on a vapor cell, and $n=100$, which is used when performing quasi-DC E-field sensing.

The Rb vapor cell is housed in a 3D-printed plastic structure on a low-profile 5-axis positioner mount. The 5-axis positioner allows the cell to be moved in relation to the overlapped lasers. The plastic cell holder has two compartments. The upper volume enables the cell body to be heated in its own chamber, with a hole in the bottom for the cell stem (cold finger) to be in a separate volume that has air blowing over it at all times. The cell stem remains colder than the rest of the cell body to control the vapor pressure, even when the upper body is not heated. Electric-field plates, positioned in parallel with the laser beam path, are implemented on the plastic structure to produce electric fields permeating through thermal insulators and the vapor cell, as conceptually illustrated in Fig.~\ref{fig:setup}. These electric-field plates can be used to generate external electric fields up to about 50 V/m for different purposes. For E-field screening characterization, we drive the plates using a square waveform at frequencies from sub-Hz to hundreds of Hz with an amplitude of a few hundred millivolts to several volts. For E-field sensitivity characterization, we drive the plates with a square waveform at frequencies from 1 kHz to greater than 10 kHz with an amplitude at tens to hundreds of millivolts to realize the concept of internal E-field biasing using an external switching E-field, and we add a reference E-field signal with a sinusoidal waveform on the plates at a few Hz with a few mV amplitude. The bias magnetic field along the direction of the laser beams is produced by a pair of B-field coils to define the quantization axis for atomic transitions via laser fields. The maximum B-field strength is about 30 G.

\begin{figure}[t]
 \begin{centering}
 	  \includegraphics[width=0.5\textwidth]{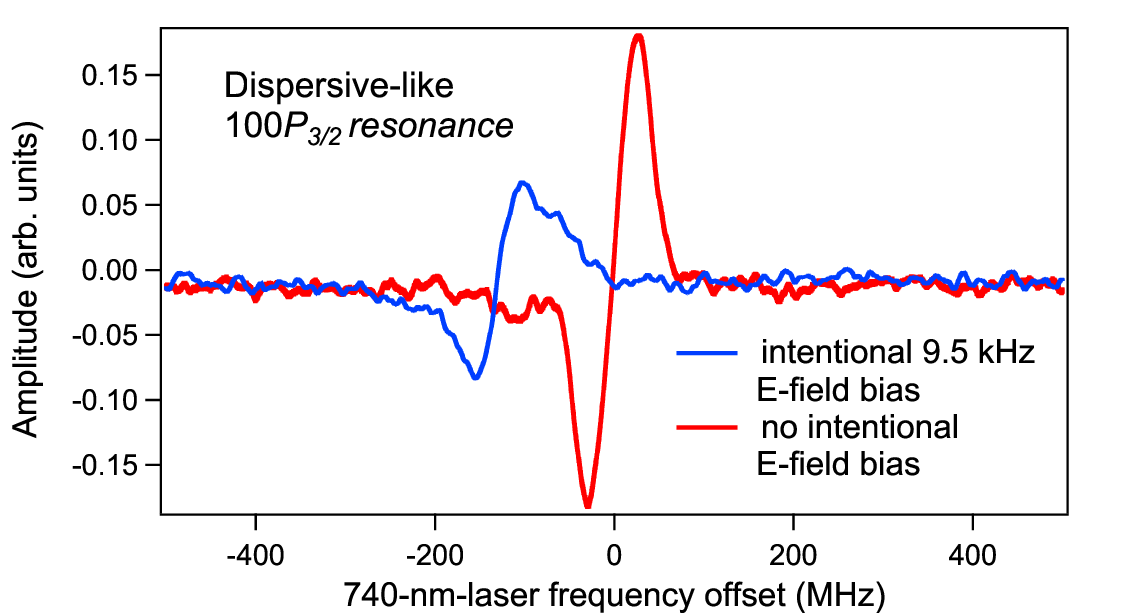}\\

  \caption{The Rydberg $100P_{3/2}$ resonance with and without internal E-field biasing. Red trace: without the 9.5 kHz external switching E-field. Blue trace: the shifted resonance with the 9.5 kHz external switching E-field. A sapphire vapor cell was used for the data. }
        \label{fig:95kHzpeakshift}
 \end{centering}
\end{figure}

For detecting the Rydberg resonance signal, the two 780-nm beams are received at the BPD after passing through the Rb vapor cell. The BPD voltage signal is sent to a lock-in amplifier (the first lock-in amplifier). This lock-in amplifier sources a frequency ranging from 10 to 15 kHz, which is sent to dither the 740 nm laser optical frequency, and demodulates the BPD voltage to generate a dispersive-like Rydberg resonance by sweeping the 740-nm laser frequency, as shown by the red trace in Fig.~\ref{fig:95kHzpeakshift}. When applying the E-field biasing inside the cell via an externally switched E-field, we observe a frequency shift of the resonance as the blue trace shown in Fig.~\ref{fig:95kHzpeakshift}. The optimal dithering frequency to the 740-nm laser is chosen to attain the best SNR of the $nP_{3/2}$ Rydberg resonance. Additionally, the dithering of the laser leads to a zero crossing of the dispersive-like resonance, as seen in Fig.~\ref{fig:95kHzpeakshift}, that can be used as an error signal to slowly lock the 740 nm laser to this Rydberg excitation transition frequency for the $nP_{3/2}$ state. Then if a changing E-field is introduced that shifts the resonance frequency faster than the laser lock loop time constant (typically $\geq0.5$ s), one can observe the change in the E-field amplitude by the changing voltage signal at the output of the lock-in amplifier.

\section{Characterization of E-field screening in vapor cells}
\begin{figure}[t]
 \begin{centering}
 	  \includegraphics[width=0.5\textwidth]{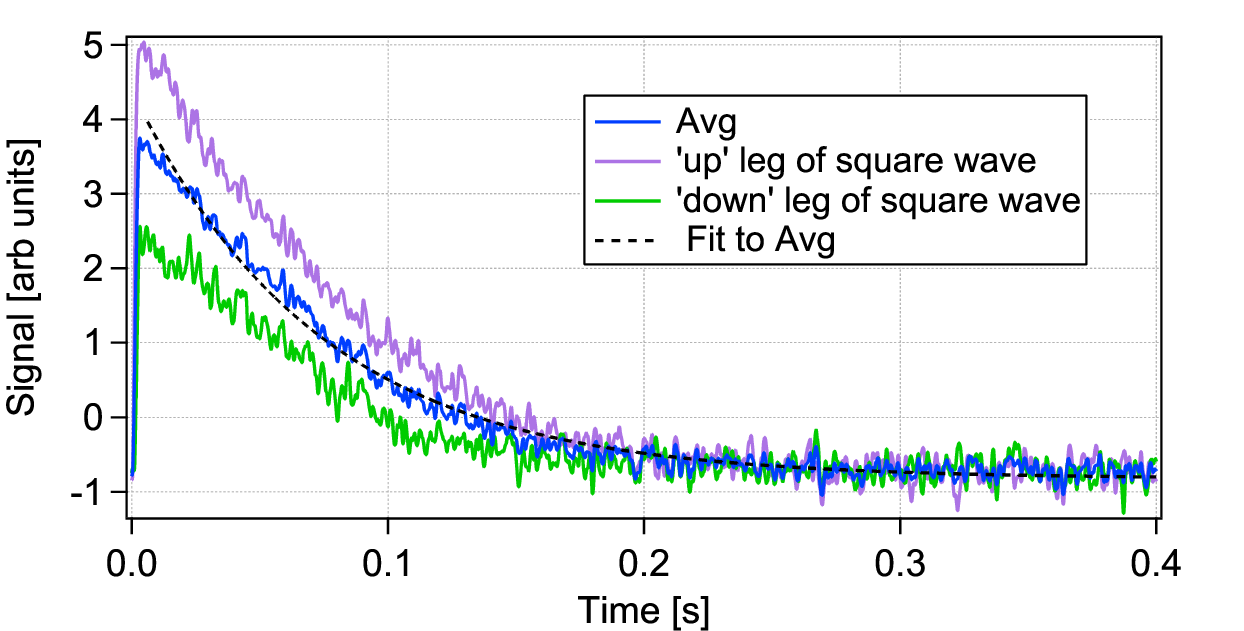}\\

  \caption{An example of the decay from a square wave, showing the two legs of the square wave can give different decays due to an internal E-field bias that can be averaged away. A sapphire vapor cell was used for the data.}
        \label{fig:decay}
 \end{centering}
\end{figure}

To characterize the E-field screening with different vapor cell coatings, we take advantage of the $50P_{3/2}$ Rydberg state because of its relatively higher transition oscillator strength, which provides sufficient SNR for a wider testing range of laser power with enough signal to measure the E-field screening rates. We characterize six different coatings on quartz vapor cells, Al$_2$O$_3$, DLC, MgO, MgAl$_2$O$_4$, paraffin, and OTS, along with two sapphire cells. The monocrystalline sapphire cell, and cells with OTS coating and paraffin coating have a natural abundance of rubidium as the fill of the vapor cell while the rest of the coated cells are filled with isotopically enriched $^{85}$Rb. Similar to the method used in Ref.~\cite{Jau2020}, a LabVIEW program is used with the lock-in amplifier output signal to lock to the center of the 740 error signal (dispersive-like Rydberg resonance) with only the integral gain to keep it on resonance for the long term. We can then apply a square wave with 50\% duty cycle to the electric-field plates and observe the impact of the fast change in electric field on the vapor cell. The center frequency of the Rydberg resonance shifts with the fast change in the electric field from the two half cycles (up and down legs) of the square wave then decaying to the original center frequency or voltage as the free charges on the inner cell move to screen the suddenly changed static electric field. The frequency of the square wave is chosen such that the whole decay trace is captured in the period. Figure~\ref{fig:decay} illustrates an example of signal traces from the  E-field screening rate measurement. The up and down legs of the square wave can end up having different heights of the decay if there are some residual internal electric fields in the vapor cell, which are commonly observed. Since the internal electric field is constant while the square wave is flipping signs for each leg, we can take an average of the higher and lower voltage signals of the square wave to obtain the effect from the square wave alone and average out the internal electric field. It is then fitted to an exponential decay function, $\exp(-2\gamma t)$, where the factor of two comes from the quadratic dependence of the electric field. As the time it takes to decay changes and/or the signal size changes, the settings for the lock-in amplifier time constant and gain, the frequency of the square wave, and the number of averages must also be adjusted to maintain similar SNR for different laser-power settings or vapor cell coatings.

\begin{figure}[t]
 \begin{centering}
 	  \includegraphics[width=0.5\textwidth]{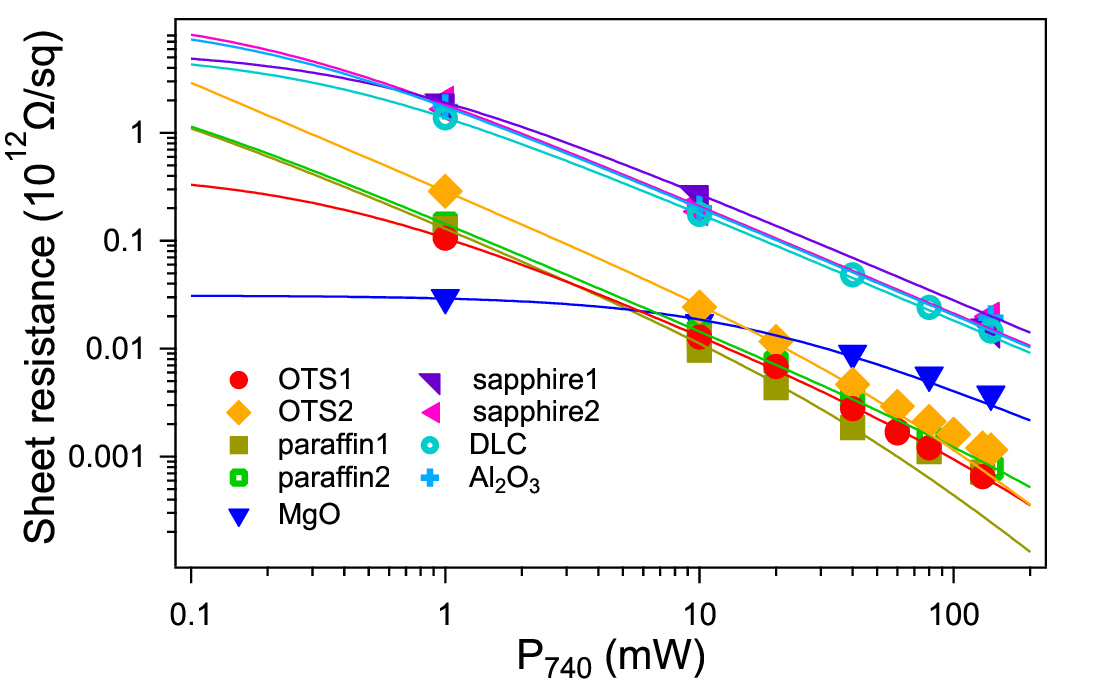}\\

  \caption{The sheet resistance $R_\Box$ of different inner-surface material versus the 740 nm laser power. The lines are fits of the associated data to 1/($K_0+K_1P_{740}+K_2P_{740}^2$) or 1/($K_0+K_1P_{740}$). The 1367 nm laser is at 100 uW, the 780 nm laser is at 200 uW, and the 795 nm laser is at 2 mW. The cell body temperature is at room temp at about 25 $^\circ$C. The Rb number density $\approx1.1\times10^{10}$ cm$^{-3}$ (measured). The magnetic field is about 9 G.}
        \label{fig:740resistance}
 \end{centering}
\end{figure}

\begin{table*}[t]
    \centering
    \footnotesize
    \begin{tabular}{|c||c|c|c|c|c|c|c|c|c|}
     \hline
        Sheet conductance$\backslash$Coating&  Al$_2$O$_3$&  DLC&  MgO&  OTS1&  OTS2&  paraffin1&  paraffin2&  sapphire1& sapphire2\\
                \hline
                        \hline
         $K_2$ ($1/{\rm T}\Omega\cdot{\rm mW^2}$)&  &  &  &  0.04 ± 0.01&  0.05 ± 0.02&   0.2 ± 0.2&   0.01± 0.01&  &

\\
         $K_1$ ($1/{\rm T}\Omega\cdot{\rm mW}$)&  0.41 ± 0.01&         0.51 ± 0.01&  1.9 ± 0.1&  7.1 ± 0.3&  3.4± 0.2&    8 ± 2&  6.8 ± 0.1&  0.35 ± 0.01&         0.46 ± 0.01\\
         $K_0$ ($1/{\rm T}\Omega$)&  0.22 ± 0.06&  0.24 ± 0.05&  33 ± 1&  2.3 ± 0.3&  0 ± 0.2&  0.2± 2&     0.2± 0.1&  0.17 ± 0.01&   0.06 ± 0.01
  \\
          \hline
    \end{tabular}
    \caption{The sheet resistance is fitted to 1/($K_0+K_1P_{740}+K_2P_{740}^2$) for OTS and paraffin and for the others it is 1/($K_0+K_1P_{740}$), where $P_{740}$ is the 740-nm laser power. Here, ${\rm T}\Omega=10^{12}\Omega$.}
    \label{tab:coatslope}
\end{table*}
Here, we do not study the dependence of the E-field screening on the cell-body temperature in depth, as described in our previous work~\cite{Jau2020}. Instead, we use the same temperature for all measurements. We investigate all of the coatings mentioned earlier with a small subset of magnetic field changes and laser power changes, with a main focus on the sapphire cell and the Al$_2$O$_3$ cell for the bulk of the experiment over a wide array of different laser power combinations of the four lasers in the system. For the sapphire and Al$_2$O$_3$ cell, we explore three different powers for each laser in all 81 variations of each laser power. However, for some of the combinations of the four lasers where all of the lasers were at low power, we were unable to obtain data as the SNR of the signal was too small to lock the 740 nm laser to the $6S_{1/2}$--$nP_{3/2}$ transition frequency. We found that the sapphire and Al$_2$O$_3$ cells behaved similarly with all of the 81 variations. Increasing the 740 nm laser power causes similar effects to the 480 nm laser from~Section V of Ref.~\cite{Jau2020}, with all of the different combinations of the other three lasers powers. For the other three lasers, once the laser power starts saturating the transition, the laser-power dependent decay rate stops being linear and starts not changing as quickly until it levels off. To eliminate the influence of cell material and geometry on the screening decay rate, once again, we use FEM method to convert the measured decay rates to the sheet resistance values $R_\Box$ for comparing the E-field screening effects with the three-photon approach. An example of changing the 740 nm laser power from 1 mW up to 140 mW, while all of the other parameters are constant, can be seen in Fig.~\ref{fig:740resistance}. Some quantitative results can be found in Table~\ref{tab:coatslope}. In summary, for minimizing the laser power dependence of the E-field screening rate, sapphire cells and glass cells with Al$_2$O$_3$ and DLC coatings are still the ``best'' options with the three-photon scheme. We expected that the paraffin and OTS-coated cells could perform better in the three-photon experiment compared to the two-photon scheme, but it does not seem to be the case. There is a possibility that these long-chain hydrocarbon-coated cells degrade over the years since the cells were purchased.
\begin{figure}[t]
 \begin{centering}
 	  \includegraphics[width=0.5\textwidth]{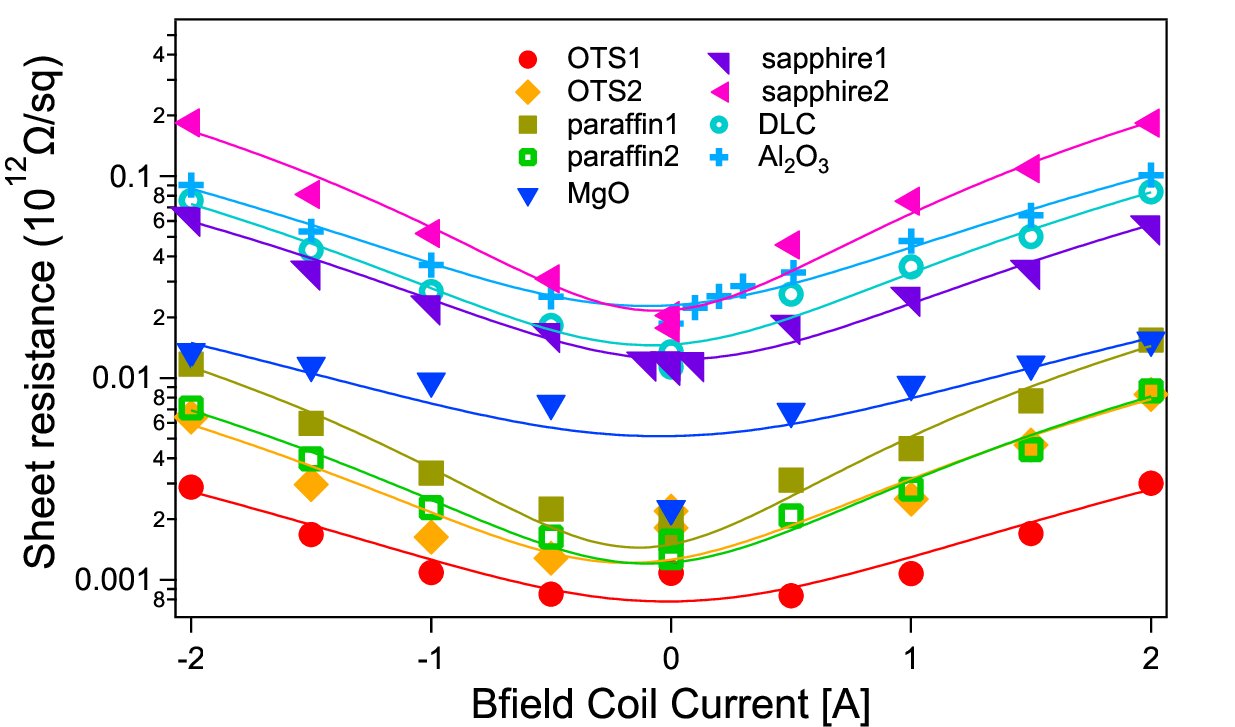}\\

  \caption{The lines are quadratic fits to the associated data. The 740 nm laser is at 80 mW, the 1367 nm laser is at 100 uW, the 780 nm laser is at 200 uW, and the 795 nm laser is at 1 mW. The cell body temperature is at room temp at about 25 $^\circ$C. The B-field coil current to gauss is 15 G/A.}
        \label{fig:bfieldcoat}
 \end{centering}
\end{figure}

\begin{figure}[t]
 \begin{centering}
 	  \includegraphics[width=0.45\textwidth]{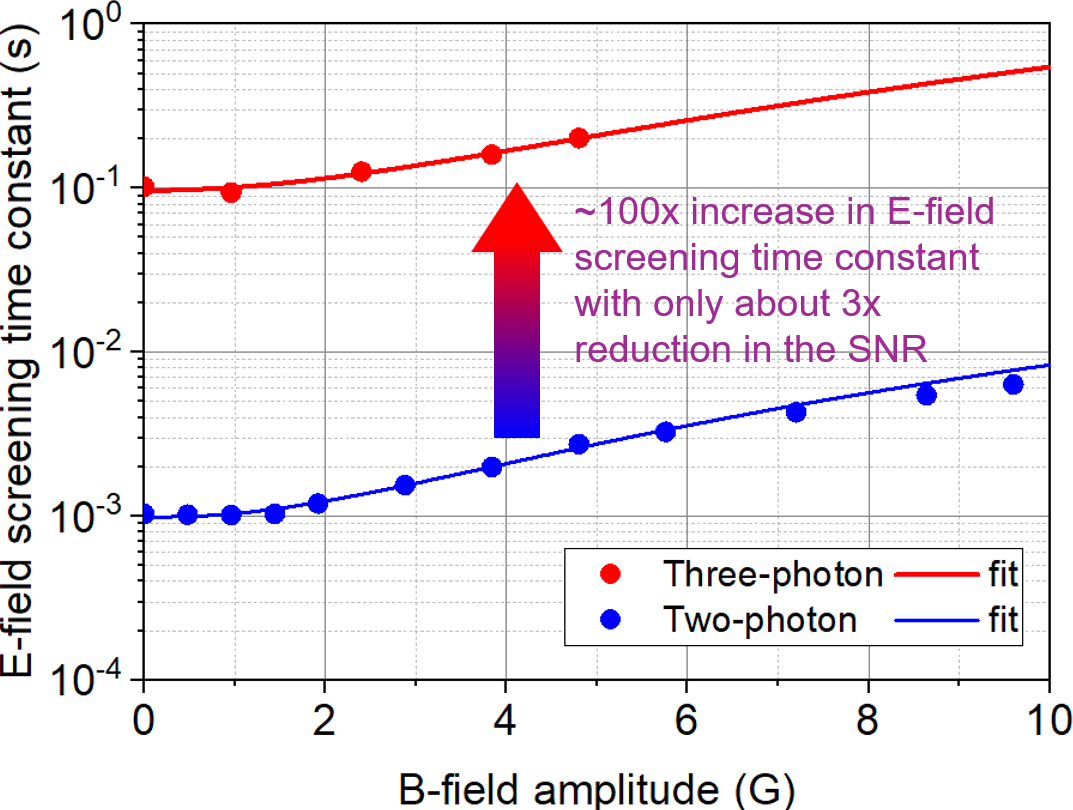}\\

  \caption{The B-field suppressed E-field screening rates using the two-photon and three-photon schemes with an Al$_2$O$_3$-coated cell. Two-photon laser power: 10 mW for 480 nm, 25 uW for 780 nm, and 1 mW for the 795 nm repump. Three-photon laser power: 10 mW for 740 nm, 20 uW for 780 nm, 100 uW for 1367 nm, and 2 mW for the 795 repump. The cell body is at about 41 $^\circ$C for both. }
        \label{fig:3v2screen}
 \end{centering}
\end{figure}

We have also experimentally characterized the B-field suppressed E-field screening effect with the three-photon scheme, as shown in Fig.~\ref{fig:bfieldcoat}. The stronger the magnetic field is, the longer it takes to decay and the larger $R_\Box$ is, for all six listed surface materials. The shape is generally quadratic with the center of the quadratic near zero B-field coil current. The quadratic behavior may be partially caused by the background magnetic field that has B-field components, which we do not compensate actively, orthogonal to the bias B-field direction. For comparing the three-photon scheme with the two photon scheme, we use the Al$_2$O$_3$-coated cell with some laser power conditions that have the 740-nm laser power the same as the 480-nm laser power used in the two-photon case. We find that the E-field screening rate is reduced by a factor of $\sim100$ with only about three times reduction in the SNR as indicated in Fig.~\ref{fig:3v2screen}. We also carry out density-matrix modeling~\cite{Happer2010} work to verify the size of signals in the two-photon and three-photon scenarios using the laser power parameters for Fig.~\ref{fig:3v2screen}.

\section{Performance of quasi-DC E-field sensing}
We exploit the E-field screening characterization with the $50P_{3/2}$ state to identify the optimal settings of the system for the starting point in obtaining better quasi-DC E-field sensing. The 740 nm laser was then tuned for the $100P_{3/2}$ state to take advantage of the stronger E-field-induced frequency shift at the higher $nP$ level. We find the best conditions for optimal E-field sensing with the vapor cell using either the sapphire cell or one of the Al$_2$O$_3$-coated and DLC-coated cells. We verify different combinations of laser power for maximizing a figure of merit (FOM), which is defined as the SNR/(resonance linewidth)$\times\kappa/\gamma$. We found that maximal power of 140 mW for the 740-nm laser power and 2 mW for the 795-nm laser power and saturating SNR powers for 780 nm at 100--200 uW and for 1367 nm at 400--600 uW are preferable to obtain the best signal. We try to use the largest magnetic field without significantly heating up the B-field coil wires, along with the largest, workable external switching E-field amplitude to attain the highest sensitivity to small changes in the detected electric field. We found that the exact positioning of the four lasers relative to each other within the beam overlap and relative to the position/angle of the vapor cell affected the E-field sensing. We optimized the overlap of the four laser beams and their positions to obtain the optimal E-field sensing.

Compared to the LED induced E-field biasing, we have verified that the E-field biasing via an external switching E-field with the second lock-in amplifier offers better E-field sensitivity, although it is still far from the ideal performance. For detecting quasi-DC electric fields, we first add a switching voltage signal to the electric-field plates. As stated in the previous sections, when applying an external switching E-field with a high enough frequency to overcome the screening from the rubidium surface adsorption, the Rydberg resonance shifts in frequency, seen in Fig.~\ref{fig:100pvss} and Fig.~\ref{fig:95kHzpeakshift}. Then similarly to the E-field screen characterization work, we use a LabVIEW program that takes the output signal from the 1st lock-in amplifier to lock the 740-nm laser to the center of the $100P_{3/2}$ resonance with a very slow lock-loop time. We then purposely introduce a small 5-Hz reference electric field and to maximize the detection of this reference signal from the output of the 2nd lock-in amplifier. In our setup, an external switching E-field above 1 kHz is fast enough to overcome the vapor-cell E-field screening. To simplify the tabletop experiment, we combine a sinusoidal reference signal with the kHz switching signal to drive the electric-field plates. The switching E-field source can also be set up independently, which is exactly what we implemented in our handheld atomic E-field sensor.

\begin{figure}[t]
 \begin{centering}
 	  \includegraphics[width=0.45\textwidth]{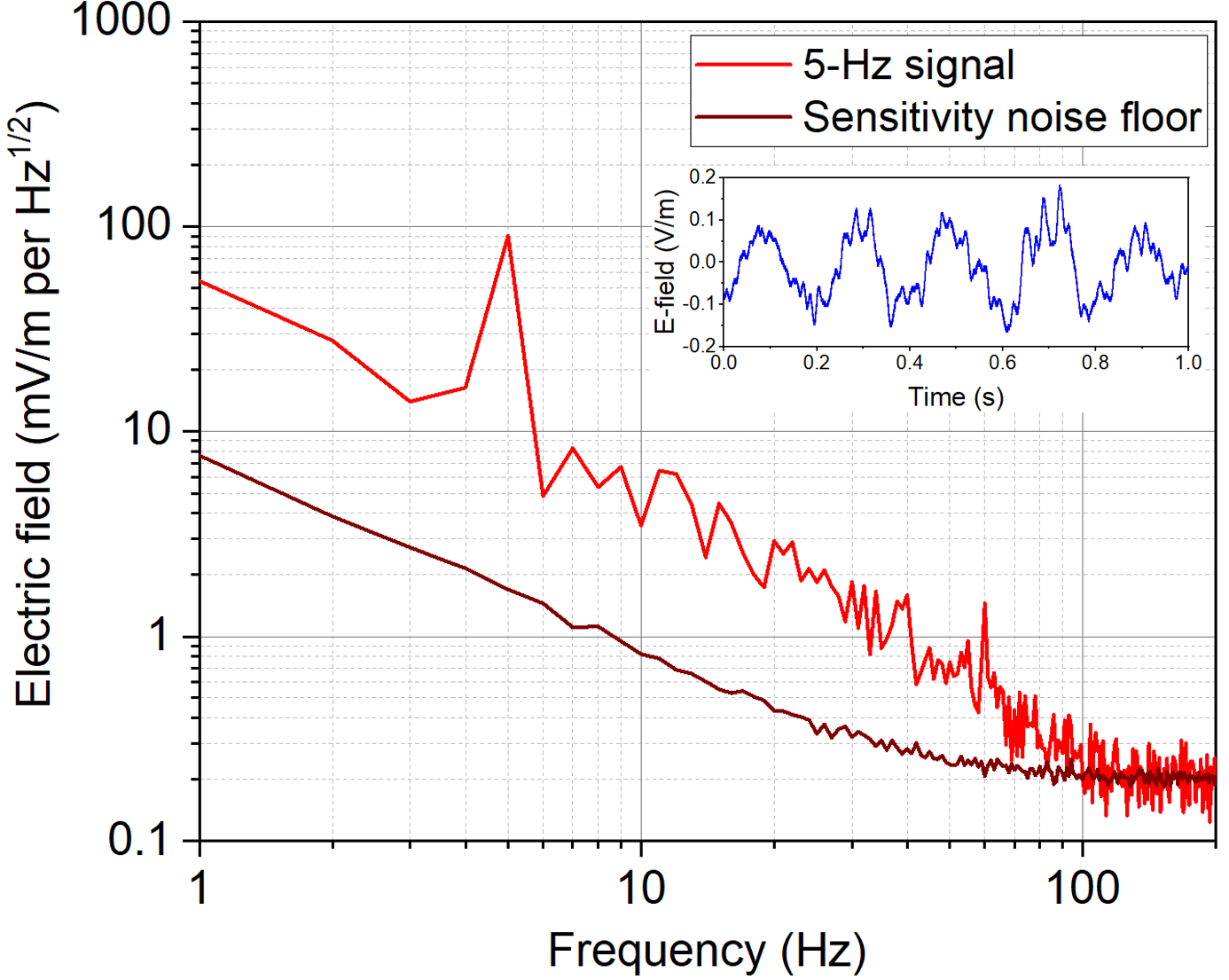}\\

  \caption{Spectrum of the 5-Hz reference signal and the sensitivity noise floor with a sapphire vapor cell. The inset is an example of the time trace of the detected 5-Hz electric field with ambient E-field noise.}
        \label{fig:5HzSpectrum}
 \end{centering}
\end{figure}
After optimizing the operating parameters with a sapphire cell, the spectrum of the detected 5-Hz reference signal and the detection noise floor are plotted in Fig.~\ref{fig:5HzSpectrum}. The driving amplitude of the 5-Hz sine wave on the electric-field plates is 5 mV, and the switching frequency is optimized with an optimal amplitude on the electric-field plates to achieve the best sensitivity. The exact electric-field amplitude at the laser beams inside the vapor cell can be obtained using an FEM simulation by considering the dimension of the electric-field plates, the nearby structures, objects, and their materials. In the quasi-DC E-field sensing experiment, we acquire the signal traces in time from the output of the 2nd lock-in amplifier with one-second time span and perform a Fourier transform to obtain the spectral information. Therefore, the spectral amplitude is naturally determined with a unit of per Hz$^{1/2}$. To minimize the random fluctuation on the spectral curves, we average power spectrum data several times, which does not impact the actual spectral amplitude but only increases the accuracy of determining the spectral SNR. In Fig.~\ref{fig:5HzSpectrum}, the 5-Hz reference E-field signal is obtained with a spectral power average of five. The noise floor in Fig.~\ref{fig:5HzSpectrum} uses a spectral power average of 50 and is determined by blocking the 740-nm laser or turning off the externally switched E-field. This confirms that the influence of the laser frequency noise on the detection noise floor is insignificant. All the spectral curves are normalized by dividing the high-pass E-field screening response~\cite{Jau2020}, which has a $f_{\rm 3 dB}\approx41$ Hz in this case. The inset in Fig.~\ref{fig:5HzSpectrum} illustrates the single-shot time trace of the detected 5-Hz electric field with ambient E-field noise.

\begin{figure}[t]
 \begin{centering}
 	  \includegraphics[width=0.4\textwidth]{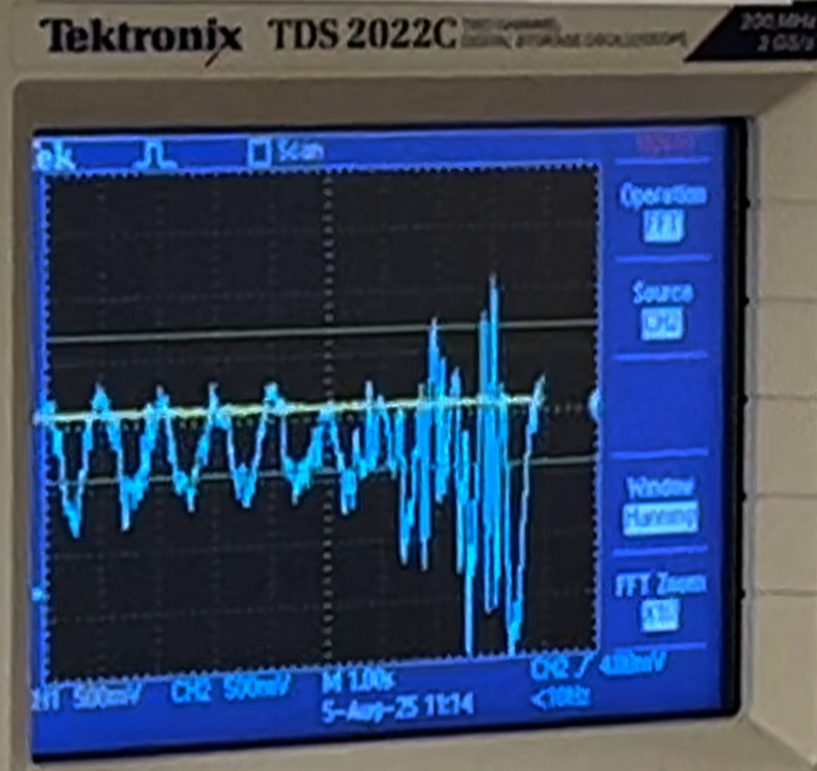}\\

  \caption{Time trace of the 1-Hz E-field signal on a oscilloscope, which gets perturbed by the noise of statics when a human finger touches a plastic object on the optical table. Please see the supplemental material for a full video.~\cite{Video}}
        \label{fig:1HzSignal}
 \end{centering}
\end{figure}
We suspect that the ambient noise associated with the 5-Hz reference signal in the spectrum (Fig.~\ref{fig:5HzSpectrum}) primarily comes from randomly occurring static noise due to the very dry laboratory environment. The quasi-DC atomic electrometry experiment presented here demonstrates superior E-field sensitivity; therefore, it is easily perturbed by various sources of noise, such as random static charges from adjacent objects due to air flowing over their surfaces or physical contact between different materials, movements of human bodies in the nearby area, and unidentified E-field noise and RF signals with amplitude noise within our detection band originating outside our laboratory space.

To exhibit how tiny static charges can impact the detected E-field signal, Figure~\ref{fig:1HzSignal} shows a picture of a detected 1-Hz E-field signal on the screen of an oscilloscope. Its E-field amplitude is about 0.45 V/m with a 25 mV driving amplitude on the electric-field plates. When a plastic object on the same optical table as the three-photon electrometry setup was gently touched by a human finger, the signal was completely scrambled. Due to the high quasi-DC E-field sensitivity of our setup, data acquisition for E-field sensing characterization must be performed in an environment with a low 60-Hz background and during ``quiet'' moments—when there are no nearby physical activities, no sudden changes in air currents within the lab, no fluctuations in room illumination, and no electromagnetic interference from adjacent labs.

In addition to using the sapphire-made Rb vapor cell, we also investigate the Al$_2$O$_3$-coated cell for E-field sensing. We find the sensitivity is also impressive, only less than a factor of two worse than a sapphire cell. According to Fig.~\ref{fig:740resistance} and Fig.~\ref{fig:bfieldcoat}, Al$_2$O$_3$-coated and DLC-coated cells are indeed good options for replacing the sapphire cells.

\section{Discussion}
In this paper, we present some scientific outcomes from our research projects over the past five years. We describe several technical schemes to further improve the E-field sensitivity in the quasi-DC frequency domain. We report our experimental studies on cell materials and cell coatings for quasi-DC E-field sensing with two-photon and three-photon Rydberg interrogations. We demonstrate significantly improved performance of quasi-DC atomic electrometry using bare vapor cells. Our work delivers a decent sensitivity level ranging from 0.2 to 7.7 mV/m$\sqrt{\rm Hz}$ for detection frequencies from 1 to 100 Hz. For the ELF band (3--30 Hz), the detection noise floor is from about 0.3 to 3 mV/m per Hz$^{1/2}$, as shown in Fig.~\ref{fig:5HzSpectrum}. Considering the nominal beam diameter of about 1 mm and the beam path length of 14 mm, the active volume (sensing volume) is about 11 mm$^3$. A small sensing volume is advantageous for detecting localized, concentrated E-field sources at short distances, which an E-field sensor with a large antenna cannot efficiently detect because the electric field decays with distance from the source faster than the gain achieved by increasing the antenna length. A large antenna is beneficial only when the E-field strength is relatively uniform over an area that is larger than the antenna dimensions.

\begin{figure}[b]
 \begin{centering}
 	  \includegraphics[width=0.48\textwidth]{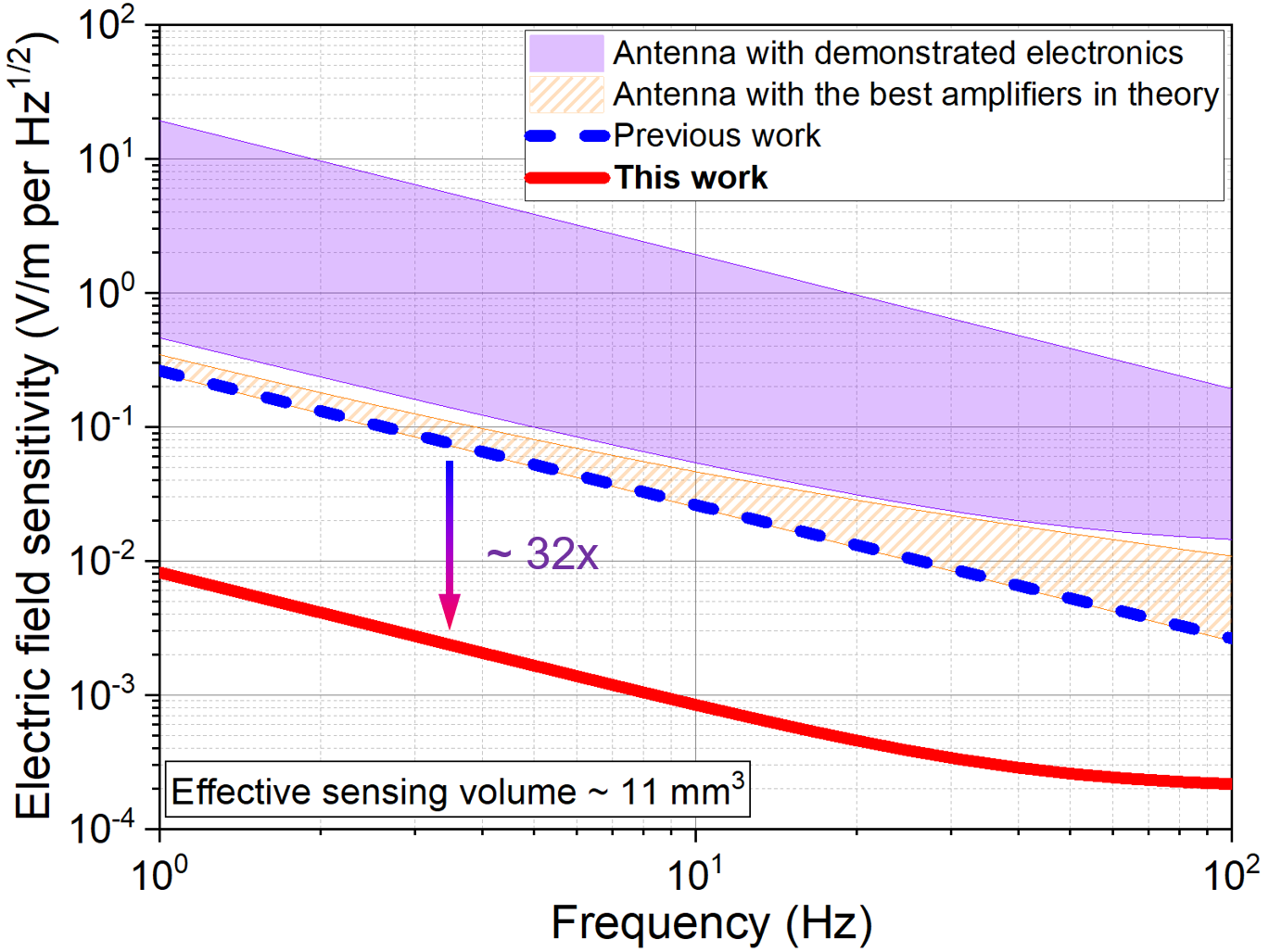}\\

  \caption{Comparison of E-field sensitivity between 1 and 100 Hz using electronic and atomic-vapor-cell approaches with about the same effective sensing volume of 11 mm$^3$. The vapor-cell sensors demonstrate the best sensitivity and the current work is about 32 times better than the previous work~\cite{Jau2020}.}
        \label{fig:AtomicVsElectronic}
 \end{centering}
\end{figure}
To have a better assessment of the performance of our vapor-cell quasi-DC E-field sensor, we compare quasi-DC E-field sensitivity using electronic and vapor-cell approaches with the same sensing volume in Fig.~\ref{fig:AtomicVsElectronic}. We established a mathematical model~\cite{ElectronicModel} to calculate the spectral E-field sensitivity of an electronic E-field detector using a dipole antenna based on the antenna dimensions and the specifications of the front-end amplification electronics. To test our electronic model, we found two separate electronic E-field sensor efforts that used large antennas and published in 2001~\cite{Krupka2001} and in 2021~\cite{Gurses2021}.The relevant parameters of the antennas and pre-amplifiers used in these two publications allow us to incorporate into the model, and the modeling results agree well with their experimental data. We can then calculate the sensitivity with a small antenna to have the same effective sensing volume as we use in the sapphire vapor cells. In Fig.~\ref{fig:AtomicVsElectronic}, the purple shaded area indicates the results of using parameters of the demonstrated electronics in the two publications; the orange striped area predicts the sensitivity using the high-end amplifier chips, such as INA116 and ADA4530; the blue-dashed line is the result from the previous work~\cite{Jau2020}; and the red-solid curve is the demonstrated sensitivity in the current work. We see a factor of about 32 improvement, which is also much better than the electronic approach.

\begin{figure}[t]
 \begin{centering}
 	  \includegraphics[width=0.48\textwidth]{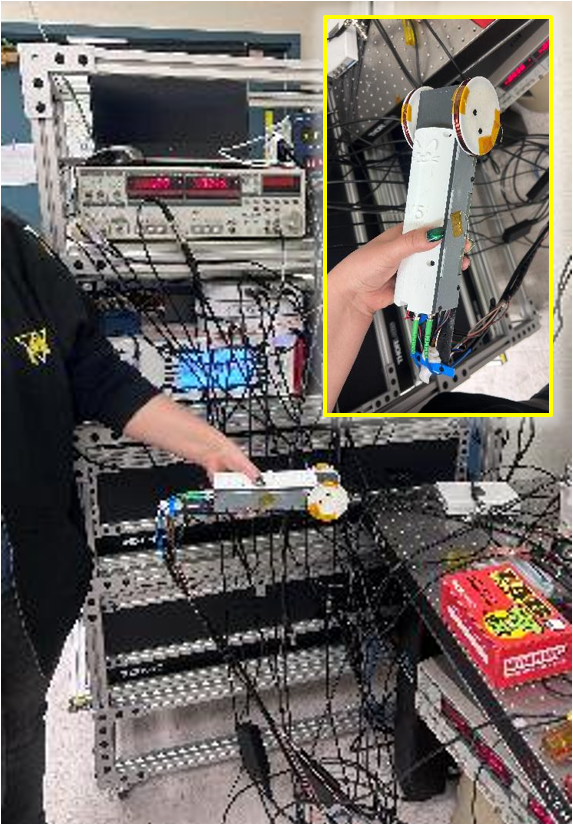}\\

  \caption{A picture of our transportable quasi-DC atomic E-field sensor testbed, featuring a handheld vapor-cell sensor head in the inset photo.}
        \label{fig:Esensor}
 \end{centering}
\end{figure}

While we show that E-field sensing in the quasi-DC realm using bare vapor cells can be continuously improved, we have not yet attained the best expected performance due to non-ideal implementation and technical imperfections, which we will strive to overcome in the future. During the preparation of this paper, we were excited to learn about another experimental demonstration of highly sensitive quasi-DC E-field detection using a miniature atomic beam with three-photon Rydberg excitation and ion counting detection inside a vacuum chamber without implementing internal electrodes~\cite{ARL2026}. During the public release process for this paper, we were also thrilled to see two new arXiv papers reporting E-field detections down to 1 kHz~\cite{Kayim2026} and to 1 Hz~\cite{Chandra2026} with bare vapor cells. In general, atomic metrology can outperform other sensing techniques when a small-form-factor sensor for low detection frequency or long-term measurement is demanded. Vapor-cell-based atomic devices also offer the best portability. We believe that advancing the performance of the quasi-DC atomic vapor-cell electrometer will enable handheld diagnostic tools for electronics without physical contacts, small-form-factor E-field-based communication devices from the SLF band to below the ELF band, and miniature sensors for proximity detection, remote activity surveillance, tracing charge signatures, and research in bioscience and geoscience. For example, in Fig.~\ref{fig:Esensor}, we show a transportable testbed with a handheld atomic E-field sensor head developed at SNL.

\begin{acknowledgments}
We would like to thank Justin Christensen, Coleman Cariker, Bjorn Sumner, and Adrian Orozco for their contributions to the experimental efforts. We thank Neil Claussen for proofreading this paper. This work was supported by the DOE NNSA. Sandia is a multimission laboratory managed and operated by National Technology and Engineering Solutions of Sandia (NTESS), LLC, a wholly owned subsidiary of Honeywell International Inc., for the U.S. Department of Energy National Nuclear Security Administration under contract DE-NA0003525. This paper describes objective technical results and analysis. Any subjective views or opinions that might be expressed in the paper do not necessarily represent the views of the U.S. Department of Energy or the United States Government.
\end{acknowledgments}

\bibliography{TPQDCAESRefs_arXiv}

\begin{thebibliography}{49}%
\makeatletter
\providecommand \@ifxundefined [1]{%
 \@ifx{#1\undefined}
}%
\providecommand \@ifnum [1]{%
 \ifnum #1\expandafter \@firstoftwo
 \else \expandafter \@secondoftwo
 \fi
}%
\providecommand \@ifx [1]{%
 \ifx #1\expandafter \@firstoftwo
 \else \expandafter \@secondoftwo
 \fi
}%
\providecommand \natexlab [1]{#1}%
\providecommand \enquote  [1]{``#1''}%
\providecommand \bibnamefont  [1]{#1}%
\providecommand \bibfnamefont [1]{#1}%
\providecommand \citenamefont [1]{#1}%
\providecommand \href@noop [0]{\@secondoftwo}%
\providecommand \href [0]{\begingroup \@sanitize@url \@href}%
\providecommand \@href[1]{\@@startlink{#1}\@@href}%
\providecommand \@@href[1]{\endgroup#1\@@endlink}%
\providecommand \@sanitize@url [0]{\catcode `\\12\catcode `\$12\catcode
  `\&12\catcode `\#12\catcode `\^12\catcode `\_12\catcode `\%12\relax}%
\providecommand \@@startlink[1]{}%
\providecommand \@@endlink[0]{}%
\providecommand \url  [0]{\begingroup\@sanitize@url \@url }%
\providecommand \@url [1]{\endgroup\@href {#1}{\urlprefix }}%
\providecommand \urlprefix  [0]{URL }%
\providecommand \Eprint [0]{\href }%
\providecommand \doibase [0]{https://doi.org/}%
\providecommand \selectlanguage [0]{\@gobble}%
\providecommand \bibinfo  [0]{\@secondoftwo}%
\providecommand \bibfield  [0]{\@secondoftwo}%
\providecommand \translation [1]{[#1]}%
\providecommand \BibitemOpen [0]{}%
\providecommand \bibitemStop [0]{}%
\providecommand \bibitemNoStop [0]{.\EOS\space}%
\providecommand \EOS [0]{\spacefactor3000\relax}%
\providecommand \BibitemShut  [1]{\csname bibitem#1\endcsname}%
\let\auto@bib@innerbib\@empty
\bibitem [{\citenamefont {Holloway}\ \emph {et~al.}(2014)\citenamefont
  {Holloway}, \citenamefont {Gordon}, \citenamefont {Jefferts}, \citenamefont
  {Schwarzkopf}, \citenamefont {Anderson}, \citenamefont {Miller},
  \citenamefont {Thaicharoen},\ and\ \citenamefont
  {Raithel}}]{Holloway2014IEEE}%
  \BibitemOpen
  \bibfield  {author} {\bibinfo {author} {\bibfnamefont {C.~L.}\ \bibnamefont
  {Holloway}}, \bibinfo {author} {\bibfnamefont {J.~A.}\ \bibnamefont
  {Gordon}}, \bibinfo {author} {\bibfnamefont {S.}~\bibnamefont {Jefferts}},
  \bibinfo {author} {\bibfnamefont {A.}~\bibnamefont {Schwarzkopf}}, \bibinfo
  {author} {\bibfnamefont {D.~A.}\ \bibnamefont {Anderson}}, \bibinfo {author}
  {\bibfnamefont {S.~A.}\ \bibnamefont {Miller}}, \bibinfo {author}
  {\bibfnamefont {N.}~\bibnamefont {Thaicharoen}},\ and\ \bibinfo {author}
  {\bibfnamefont {G.}~\bibnamefont {Raithel}},\ }\bibfield  {title} {\bibinfo
  {title} {Broadband rydberg atom-based electric-field probe for si-traceable,
  self-calibrated measurements},\ }\href@noop {} {\bibfield  {journal}
  {\bibinfo  {journal} {IEEE Transactions on Antennas and Propagation}\
  }\textbf {\bibinfo {volume} {62}},\ \bibinfo {pages} {6169} (\bibinfo {year}
  {2014})}\BibitemShut {NoStop}%
\bibitem [{\citenamefont {Sedlacek}\ \emph {et~al.}(2012)\citenamefont
  {Sedlacek}, \citenamefont {Schwettmann}, \citenamefont {K\"{u}bler},
  \citenamefont {L\"{o}w}, \citenamefont {Pfau},\ and\ \citenamefont
  {Shaffer}}]{Sedlacek2012}%
  \BibitemOpen
  \bibfield  {author} {\bibinfo {author} {\bibfnamefont {J.~A.}\ \bibnamefont
  {Sedlacek}}, \bibinfo {author} {\bibfnamefont {A.}~\bibnamefont
  {Schwettmann}}, \bibinfo {author} {\bibfnamefont {H.}~\bibnamefont
  {K\"{u}bler}}, \bibinfo {author} {\bibfnamefont {R.}~\bibnamefont {L\"{o}w}},
  \bibinfo {author} {\bibfnamefont {T.}~\bibnamefont {Pfau}},\ and\ \bibinfo
  {author} {\bibfnamefont {J.~P.}\ \bibnamefont {Shaffer}},\ }\bibfield
  {title} {\bibinfo {title} {Microwave electrometry with {R}ydberg atoms in a
  vapour cell using bright atomic resonances},\ }\href@noop {} {\bibfield
  {journal} {\bibinfo  {journal} {Nat. Phys.}\ }\textbf {\bibinfo {volume}
  {8}},\ \bibinfo {pages} {819} (\bibinfo {year} {2012})}\BibitemShut {NoStop}%
\bibitem [{\citenamefont {Miller}\ \emph {et~al.}(2016)\citenamefont {Miller},
  \citenamefont {Anderson},\ and\ \citenamefont {Raithel}}]{Miller2016}%
  \BibitemOpen
  \bibfield  {author} {\bibinfo {author} {\bibfnamefont {S.~A.}\ \bibnamefont
  {Miller}}, \bibinfo {author} {\bibfnamefont {D.~A.}\ \bibnamefont
  {Anderson}},\ and\ \bibinfo {author} {\bibfnamefont {G.}~\bibnamefont
  {Raithel}},\ }\bibfield  {title} {\bibinfo {title} {Radio-frequency-modulated
  {R}ydberg states in a vapor cell},\ }\href@noop {} {\bibfield  {journal}
  {\bibinfo  {journal} {New J. Phys.}\ }\textbf {\bibinfo {volume} {18}},\
  \bibinfo {pages} {053017} (\bibinfo {year} {2016})}\BibitemShut {NoStop}%
\bibitem [{\citenamefont {Wade}\ \emph {et~al.}(2018)\citenamefont {Wade},
  \citenamefont {M.~Marcuzzi}, \citenamefont {Kondo}, \citenamefont
  {Lesanovsky}, \citenamefont {Adams},\ and\ \citenamefont
  {Weatherill}}]{Wade2018}%
  \BibitemOpen
  \bibfield  {author} {\bibinfo {author} {\bibfnamefont {C.~G.}\ \bibnamefont
  {Wade}}, \bibinfo {author} {\bibfnamefont {E.~L.}\ \bibnamefont
  {M.~Marcuzzi}}, \bibinfo {author} {\bibfnamefont {J.~M.}\ \bibnamefont
  {Kondo}}, \bibinfo {author} {\bibfnamefont {I.}~\bibnamefont {Lesanovsky}},
  \bibinfo {author} {\bibfnamefont {C.~S.}\ \bibnamefont {Adams}},\ and\
  \bibinfo {author} {\bibfnamefont {K.~J.}\ \bibnamefont {Weatherill}},\
  }\bibfield  {title} {\bibinfo {title} {A terahertz-driven non-equilibrium
  phase transition in a room temperature atomic vapour},\ }\href@noop {}
  {\bibfield  {journal} {\bibinfo  {journal} {Nat. Comm.}\ }\textbf {\bibinfo
  {volume} {9}},\ \bibinfo {pages} {3567} (\bibinfo {year} {2018})}\BibitemShut
  {NoStop}%
\bibitem [{\citenamefont {Cox}\ \emph {et~al.}(2018)\citenamefont {Cox},
  \citenamefont {Meyer}, \citenamefont {Fatemi},\ and\ \citenamefont
  {Kunz}}]{cox2018}%
  \BibitemOpen
  \bibfield  {author} {\bibinfo {author} {\bibfnamefont {K.~C.}\ \bibnamefont
  {Cox}}, \bibinfo {author} {\bibfnamefont {D.~H.}\ \bibnamefont {Meyer}},
  \bibinfo {author} {\bibfnamefont {F.~K.}\ \bibnamefont {Fatemi}},\ and\
  \bibinfo {author} {\bibfnamefont {P.~D.}\ \bibnamefont {Kunz}},\ }\bibfield
  {title} {\bibinfo {title} {Quantum-limited atomic receiver in the
  electrically small regime},\ }\href
  {https://doi.org/10.1103/PhysRevLett.121.110502} {\bibfield  {journal}
  {\bibinfo  {journal} {Phys. Rev. Lett.}\ }\textbf {\bibinfo {volume} {121}},\
  \bibinfo {pages} {110502} (\bibinfo {year} {2018})}\BibitemShut {NoStop}%
\bibitem [{\citenamefont {Paradis}\ \emph {et~al.}(2019)\citenamefont
  {Paradis}, \citenamefont {Raithel},\ and\ \citenamefont
  {Anderson}}]{Paradis2019}%
  \BibitemOpen
  \bibfield  {author} {\bibinfo {author} {\bibfnamefont {E.}~\bibnamefont
  {Paradis}}, \bibinfo {author} {\bibfnamefont {G.}~\bibnamefont {Raithel}},\
  and\ \bibinfo {author} {\bibfnamefont {D.~A.}\ \bibnamefont {Anderson}},\
  }\bibfield  {title} {\bibinfo {title} {Atomic measurements of high-intensity
  vhf-band radio-frequency fields with a rydberg vapor-cell detector},\
  }\href@noop {} {\bibfield  {journal} {\bibinfo  {journal} {Phys. Rev. A}\
  }\textbf {\bibinfo {volume} {100}},\ \bibinfo {pages} {013420} (\bibinfo
  {year} {2019})}\BibitemShut {NoStop}%
\bibitem [{\citenamefont {Meyer}\ \emph {et~al.}(2021)\citenamefont {Meyer},
  \citenamefont {Kunz},\ and\ \citenamefont {Cox}}]{Meyer2021}%
  \BibitemOpen
  \bibfield  {author} {\bibinfo {author} {\bibfnamefont {D.~H.}\ \bibnamefont
  {Meyer}}, \bibinfo {author} {\bibfnamefont {P.~D.}\ \bibnamefont {Kunz}},\
  and\ \bibinfo {author} {\bibfnamefont {K.~C.}\ \bibnamefont {Cox}},\
  }\bibfield  {title} {\bibinfo {title} {Waveguide-coupled rydberg spectrum
  analyzer from 0 to 20 ghz},\ }\href@noop {} {\bibfield  {journal} {\bibinfo
  {journal} {Phys. Rev. Applied}\ }\textbf {\bibinfo {volume} {15}},\ \bibinfo
  {pages} {014053} (\bibinfo {year} {2021})}\BibitemShut {NoStop}%
\bibitem [{\citenamefont {Mingyong}\ \emph {et~al.}(2020)\citenamefont
  {Mingyong}, \citenamefont {Hu}, \citenamefont {Ma}, \citenamefont {Zhang},
  \citenamefont {Linjie}, \citenamefont {Liantuan},\ and\ \citenamefont
  {Suotang}}]{mingyong2020}%
  \BibitemOpen
  \bibfield  {author} {\bibinfo {author} {\bibfnamefont {J.}~\bibnamefont
  {Mingyong}}, \bibinfo {author} {\bibfnamefont {Y.}~\bibnamefont {Hu}},
  \bibinfo {author} {\bibfnamefont {J.}~\bibnamefont {Ma}}, \bibinfo {author}
  {\bibfnamefont {H.}~\bibnamefont {Zhang}}, \bibinfo {author} {\bibfnamefont
  {Z.}~\bibnamefont {Linjie}}, \bibinfo {author} {\bibfnamefont
  {X.}~\bibnamefont {Liantuan}},\ and\ \bibinfo {author} {\bibfnamefont
  {J.}~\bibnamefont {Suotang}},\ }\bibfield  {title} {\bibinfo {title} {Atomic
  superheterodyne receiver based on microwave-dressed rydberg spectroscopy},\
  }\href
  {https://www.proquest.com/scholarly-journals/atomic-superheterodyne-receiver-based-on/docview/2440211981/se-2}
  {\bibfield  {journal} {\bibinfo  {journal} {Nature Physics}\ }\textbf
  {\bibinfo {volume} {16}},\ \bibinfo {pages} {911} (\bibinfo {year}
  {2020})}\BibitemShut {NoStop}%
\bibitem [{\citenamefont {Holloway}\ \emph {et~al.}(2021)\citenamefont
  {Holloway}, \citenamefont {Simons}, \citenamefont {Haddab}, \citenamefont
  {Gordon}, \citenamefont {Anderson}, \citenamefont {Raithel}, ,\ and\
  \citenamefont {Voran}}]{Holloway2021}%
  \BibitemOpen
  \bibfield  {author} {\bibinfo {author} {\bibfnamefont {C.~L.}\ \bibnamefont
  {Holloway}}, \bibinfo {author} {\bibfnamefont {M.~T.}\ \bibnamefont
  {Simons}}, \bibinfo {author} {\bibfnamefont {A.~H.}\ \bibnamefont {Haddab}},
  \bibinfo {author} {\bibfnamefont {J.~A.}\ \bibnamefont {Gordon}}, \bibinfo
  {author} {\bibfnamefont {D.~A.}\ \bibnamefont {Anderson}}, \bibinfo {author}
  {\bibfnamefont {G.}~\bibnamefont {Raithel}}, ,\ and\ \bibinfo {author}
  {\bibfnamefont {S.~D.}\ \bibnamefont {Voran}},\ }\bibfield  {title} {\bibinfo
  {title} {A multiple-band rydberg atom-based receiver: Am/fm stereo
  reception},\ }\href@noop {} {\bibfield  {journal} {\bibinfo  {journal} {IEEE
  Antennas and Propagation Magazine}\ }\textbf {\bibinfo {volume} {63}},\
  \bibinfo {pages} {63} (\bibinfo {year} {2021})}\BibitemShut {NoStop}%
\bibitem [{\citenamefont {Liu}\ \emph {et~al.}(2022)\citenamefont {Liu},
  \citenamefont {Zhang}, \citenamefont {Liu}, \citenamefont {Zhang},
  \citenamefont {Zhu}, \citenamefont {Gao}, \citenamefont {Guo}, \citenamefont
  {Ding},\ and\ \citenamefont {Shi}}]{bang2022}%
  \BibitemOpen
  \bibfield  {author} {\bibinfo {author} {\bibfnamefont {B.}~\bibnamefont
  {Liu}}, \bibinfo {author} {\bibfnamefont {L.-H.}\ \bibnamefont {Zhang}},
  \bibinfo {author} {\bibfnamefont {Z.-K.}\ \bibnamefont {Liu}}, \bibinfo
  {author} {\bibfnamefont {Z.-Y.}\ \bibnamefont {Zhang}}, \bibinfo {author}
  {\bibfnamefont {Z.-H.}\ \bibnamefont {Zhu}}, \bibinfo {author} {\bibfnamefont
  {W.}~\bibnamefont {Gao}}, \bibinfo {author} {\bibfnamefont {G.-C.}\
  \bibnamefont {Guo}}, \bibinfo {author} {\bibfnamefont {D.-S.}\ \bibnamefont
  {Ding}},\ and\ \bibinfo {author} {\bibfnamefont {B.-S.}\ \bibnamefont
  {Shi}},\ }\bibfield  {title} {\bibinfo {title} {Highly sensitive measurement
  of a megahertz rf electric field with a rydberg-atom sensor},\ }\href
  {https://doi.org/10.1103/PhysRevApplied.18.014045} {\bibfield  {journal}
  {\bibinfo  {journal} {Phys. Rev. Appl.}\ }\textbf {\bibinfo {volume} {18}},\
  \bibinfo {pages} {014045} (\bibinfo {year} {2022})}\BibitemShut {NoStop}%
\bibitem [{\citenamefont {Rotunno}\ \emph {et~al.}(2023)\citenamefont
  {Rotunno}, \citenamefont {Berweger}, \citenamefont {Prajapati}, \citenamefont
  {Simons}, \citenamefont {Artusio-Glimpse}, \citenamefont {Holloway},
  \citenamefont {Jayaseelan}, \citenamefont {Potvliege},\ and\ \citenamefont
  {Adams}}]{rotunno2023}%
  \BibitemOpen
  \bibfield  {author} {\bibinfo {author} {\bibfnamefont {A.~P.}\ \bibnamefont
  {Rotunno}}, \bibinfo {author} {\bibfnamefont {S.}~\bibnamefont {Berweger}},
  \bibinfo {author} {\bibfnamefont {N.}~\bibnamefont {Prajapati}}, \bibinfo
  {author} {\bibfnamefont {M.~T.}\ \bibnamefont {Simons}}, \bibinfo {author}
  {\bibfnamefont {A.~B.}\ \bibnamefont {Artusio-Glimpse}}, \bibinfo {author}
  {\bibfnamefont {C.~L.}\ \bibnamefont {Holloway}}, \bibinfo {author}
  {\bibfnamefont {M.}~\bibnamefont {Jayaseelan}}, \bibinfo {author}
  {\bibfnamefont {R.~M.}\ \bibnamefont {Potvliege}},\ and\ \bibinfo {author}
  {\bibfnamefont {C.~S.}\ \bibnamefont {Adams}},\ }\bibfield  {title} {\bibinfo
  {title} {Detection of 3–300 mhz electric fields using floquet sideband gaps
  by “rabi matching” dressed rydberg atoms},\ }\href
  {https://doi.org/10.1063/5.0162101} {\bibfield  {journal} {\bibinfo
  {journal} {Journal of Applied Physics}\ }\textbf {\bibinfo {volume} {134}},\
  \bibinfo {pages} {134501} (\bibinfo {year} {2023})}\BibitemShut {NoStop}%
\bibitem [{\citenamefont {Nowosielski}\ \emph {et~al.}(2024)\citenamefont
  {Nowosielski}, \citenamefont {Jastrzebski}, \citenamefont {Halavach},
  \citenamefont {{\L}ukanowski}, \citenamefont {Jarzyna}, \citenamefont
  {Mazelanik}, \citenamefont {Wasilewski},\ and\ \citenamefont
  {Parniak}}]{nowosielski2024}%
  \BibitemOpen
  \bibfield  {author} {\bibinfo {author} {\bibfnamefont {J.}~\bibnamefont
  {Nowosielski}}, \bibinfo {author} {\bibfnamefont {M.}~\bibnamefont
  {Jastrzebski}}, \bibinfo {author} {\bibfnamefont {P.}~\bibnamefont
  {Halavach}}, \bibinfo {author} {\bibfnamefont {K.}~\bibnamefont
  {{\L}ukanowski}}, \bibinfo {author} {\bibfnamefont {M.}~\bibnamefont
  {Jarzyna}}, \bibinfo {author} {\bibfnamefont {M.}~\bibnamefont {Mazelanik}},
  \bibinfo {author} {\bibfnamefont {W.}~\bibnamefont {Wasilewski}},\ and\
  \bibinfo {author} {\bibfnamefont {M.}~\bibnamefont {Parniak}},\ }\bibfield
  {title} {\bibinfo {title} {Warm rydberg atom-based quadrature
  amplitude-modulated receiver},\ }\href {https://doi.org/10.1364/OE.529977}
  {\bibfield  {journal} {\bibinfo  {journal} {Opt. Express}\ }\textbf {\bibinfo
  {volume} {32}},\ \bibinfo {pages} {30027} (\bibinfo {year}
  {2024})}\BibitemShut {NoStop}%
\bibitem [{\citenamefont {Yang}\ \emph {et~al.}(2024)\citenamefont {Yang},
  \citenamefont {Jing}, \citenamefont {Zhang}, \citenamefont {Zhang},
  \citenamefont {Xiao},\ and\ \citenamefont {Jia}}]{yang2024}%
  \BibitemOpen
  \bibfield  {author} {\bibinfo {author} {\bibfnamefont {W.}~\bibnamefont
  {Yang}}, \bibinfo {author} {\bibfnamefont {M.}~\bibnamefont {Jing}}, \bibinfo
  {author} {\bibfnamefont {H.}~\bibnamefont {Zhang}}, \bibinfo {author}
  {\bibfnamefont {L.}~\bibnamefont {Zhang}}, \bibinfo {author} {\bibfnamefont
  {L.}~\bibnamefont {Xiao}},\ and\ \bibinfo {author} {\bibfnamefont
  {S.}~\bibnamefont {Jia}},\ }\bibfield  {title} {\bibinfo {title} {Radio
  frequency electric field-enhanced sensing based on the rydberg atom-based
  superheterodyne receiver},\ }\href {https://doi.org/10.1364/OL.522466}
  {\bibfield  {journal} {\bibinfo  {journal} {Opt. Lett.}\ }\textbf {\bibinfo
  {volume} {49}},\ \bibinfo {pages} {2938} (\bibinfo {year}
  {2024})}\BibitemShut {NoStop}%
\bibitem [{\citenamefont {Wan}\ \emph {et~al.}(2025)\citenamefont {Wan},
  \citenamefont {Lin}, \citenamefont {Yang}, \citenamefont {Zhou},\ and\
  \citenamefont {Fu}}]{wan2025}%
  \BibitemOpen
  \bibfield  {author} {\bibinfo {author} {\bibfnamefont {W.}~\bibnamefont
  {Wan}}, \bibinfo {author} {\bibfnamefont {Y.}~\bibnamefont {Lin}}, \bibinfo
  {author} {\bibfnamefont {K.}~\bibnamefont {Yang}}, \bibinfo {author}
  {\bibfnamefont {A.}~\bibnamefont {Zhou}},\ and\ \bibinfo {author}
  {\bibfnamefont {Y.}~\bibnamefont {Fu}},\ }\bibfield  {title} {\bibinfo
  {title} {Broadband tunable ultra-compact resonator enhanced rydberg atomic
  sensor},\ }\href {https://doi.org/10.1364/OE.578039} {\bibfield  {journal}
  {\bibinfo  {journal} {Opt. Express}\ }\textbf {\bibinfo {volume} {33}},\
  \bibinfo {pages} {53538} (\bibinfo {year} {2025})}\BibitemShut {NoStop}%
\bibitem [{\citenamefont {Zhou}\ \emph {et~al.}(2025)\citenamefont {Zhou},
  \citenamefont {Lin}, \citenamefont {Mao}, \citenamefont {Yang}, \citenamefont
  {Ding}, \citenamefont {Wan},\ and\ \citenamefont {Fu}}]{zhou2025}%
  \BibitemOpen
  \bibfield  {author} {\bibinfo {author} {\bibfnamefont {A.}~\bibnamefont
  {Zhou}}, \bibinfo {author} {\bibfnamefont {Y.}~\bibnamefont {Lin}}, \bibinfo
  {author} {\bibfnamefont {R.}~\bibnamefont {Mao}}, \bibinfo {author}
  {\bibfnamefont {K.}~\bibnamefont {Yang}}, \bibinfo {author} {\bibfnamefont
  {Z.}~\bibnamefont {Ding}}, \bibinfo {author} {\bibfnamefont {W.}~\bibnamefont
  {Wan}},\ and\ \bibinfo {author} {\bibfnamefont {Y.}~\bibnamefont {Fu}},\
  }\bibfield  {title} {\bibinfo {title} {High-sensitivity rydberg atom-based
  field sensing enhancement using miniaturized resonator},\ }\href
  {https://doi.org/10.1109/TAP.2025.3596373} {\bibfield  {journal} {\bibinfo
  {journal} {IEEE TRANSACTIONS ON ANTENNAS AND PROPAGATION}\ }\textbf {\bibinfo
  {volume} {73}},\ \bibinfo {pages} {10948} (\bibinfo {year}
  {2025})}\BibitemShut {NoStop}%
\bibitem [{\citenamefont {Manchaiah}\ \emph {et~al.}(2026)\citenamefont
  {Manchaiah}, \citenamefont {Oliver}, \citenamefont {Berweger}, \citenamefont
  {Holloway},\ and\ \citenamefont {Prajapati}}]{manchaiah2026}%
  \BibitemOpen
  \bibfield  {author} {\bibinfo {author} {\bibfnamefont {D.}~\bibnamefont
  {Manchaiah}}, \bibinfo {author} {\bibfnamefont {S.}~\bibnamefont {Oliver}},
  \bibinfo {author} {\bibfnamefont {S.}~\bibnamefont {Berweger}}, \bibinfo
  {author} {\bibfnamefont {C.~L.}\ \bibnamefont {Holloway}},\ and\ \bibinfo
  {author} {\bibfnamefont {N.}~\bibnamefont {Prajapati}},\ }\bibfield  {title}
  {\bibinfo {title} {Probing bandwidth and sensitivity in rydberg atom sensing
  via optical homodyne and rf heterodyne detection},\ }\href
  {https://doi.org/10.1103/jsbl-45t9} {\bibfield  {journal} {\bibinfo
  {journal} {Phys. Rev. A}\ }\textbf {\bibinfo {volume} {113}},\ \bibinfo
  {pages} {013729} (\bibinfo {year} {2026})}\BibitemShut {NoStop}%
\bibitem [{\citenamefont {Zhang}\ \emph {et~al.}(2026)\citenamefont {Zhang},
  \citenamefont {Sun}, \citenamefont {Yao}, \citenamefont {Zhao}, \citenamefont
  {Lin}, \citenamefont {Sang}, \citenamefont {Yang}, \citenamefont {An},\ and\
  \citenamefont {Fu}}]{Zhang2026}%
  \BibitemOpen
  \bibfield  {author} {\bibinfo {author} {\bibfnamefont {J.}~\bibnamefont
  {Zhang}}, \bibinfo {author} {\bibfnamefont {Z.}~\bibnamefont {Sun}}, \bibinfo
  {author} {\bibfnamefont {J.}~\bibnamefont {Yao}}, \bibinfo {author}
  {\bibfnamefont {F.}~\bibnamefont {Zhao}}, \bibinfo {author} {\bibfnamefont
  {Y.}~\bibnamefont {Lin}}, \bibinfo {author} {\bibfnamefont {D.}~\bibnamefont
  {Sang}}, \bibinfo {author} {\bibfnamefont {K.}~\bibnamefont {Yang}}, \bibinfo
  {author} {\bibfnamefont {Q.}~\bibnamefont {An}},\ and\ \bibinfo {author}
  {\bibfnamefont {Y.}~\bibnamefont {Fu}},\ }\bibfield  {title} {\bibinfo
  {title} {Self-dressing rydberg atomic receiver based on laser-induced dc
  field},\ }\href@noop {} {\bibfield  {journal} {\bibinfo  {journal} {npj
  Quantum Mater.}\ }\textbf {\bibinfo {volume} {11}},\ \bibinfo {pages} {28}
  (\bibinfo {year} {2026})}\BibitemShut {NoStop}%
\bibitem [{Ele()}]{ElectronicRFReceiver}%
  \BibitemOpen
  \href@noop {} {}\bibinfo {note} {For example, a typical mini GPS module can
  have sensitivity about -165 dBm with about 1 cm$^2$ antenna size and 50 bps
  data rate. The equivalent sensitivity is therefore about $1\times10^{-7}$ V/m
  per Hz$^{1/2}$ or 1 nV/cm per Hz$^{1/2}$.}\BibitemShut {Stop}%
\bibitem [{\citenamefont {Jau}\ and\ \citenamefont {Carter}(2020)}]{Jau2020}%
  \BibitemOpen
  \bibfield  {author} {\bibinfo {author} {\bibfnamefont {Y.-Y.}\ \bibnamefont
  {Jau}}\ and\ \bibinfo {author} {\bibfnamefont {T.}~\bibnamefont {Carter}},\
  }\bibfield  {title} {\bibinfo {title} {Vapor-cell-based atomic electrometry
  for detection frequencies below 1 khz},\ }\href@noop {} {\bibfield  {journal}
  {\bibinfo  {journal} {Phys. Rev. Applied}\ }\textbf {\bibinfo {volume}
  {13}},\ \bibinfo {pages} {054034} (\bibinfo {year} {2020})}\BibitemShut
  {NoStop}%
\bibitem [{\citenamefont {Mohapatra}\ \emph {et~al.}(2007)\citenamefont
  {Mohapatra}, \citenamefont {Jackson},\ and\ \citenamefont
  {Adams}}]{mohapatra2007}%
  \BibitemOpen
  \bibfield  {author} {\bibinfo {author} {\bibfnamefont {A.~K.}\ \bibnamefont
  {Mohapatra}}, \bibinfo {author} {\bibfnamefont {T.~R.}\ \bibnamefont
  {Jackson}},\ and\ \bibinfo {author} {\bibfnamefont {C.~S.}\ \bibnamefont
  {Adams}},\ }\bibfield  {title} {\bibinfo {title} {Coherent optical detection
  of highly excited rydberg states using electromagnetically induced
  transparency},\ }\href {https://doi.org/10.1103/PhysRevLett.98.113003}
  {\bibfield  {journal} {\bibinfo  {journal} {Phys. Rev. Lett.}\ }\textbf
  {\bibinfo {volume} {98}},\ \bibinfo {pages} {113003} (\bibinfo {year}
  {2007})}\BibitemShut {NoStop}%
\bibitem [{\citenamefont {Holloway}\ \emph {et~al.}(2022)\citenamefont
  {Holloway}, \citenamefont {Prajapati}, \citenamefont {Sherman}, \citenamefont
  {Rüfenacht}, \citenamefont {Artusio-Glimpse}, \citenamefont {Simons},
  \citenamefont {Robinson}, \citenamefont {La~Mantia},\ and\ \citenamefont
  {Norrgard}}]{holloway2022}%
  \BibitemOpen
  \bibfield  {author} {\bibinfo {author} {\bibfnamefont {C.~L.}\ \bibnamefont
  {Holloway}}, \bibinfo {author} {\bibfnamefont {N.}~\bibnamefont {Prajapati}},
  \bibinfo {author} {\bibfnamefont {J.~A.}\ \bibnamefont {Sherman}}, \bibinfo
  {author} {\bibfnamefont {A.}~\bibnamefont {Rüfenacht}}, \bibinfo {author}
  {\bibfnamefont {A.~B.}\ \bibnamefont {Artusio-Glimpse}}, \bibinfo {author}
  {\bibfnamefont {M.~T.}\ \bibnamefont {Simons}}, \bibinfo {author}
  {\bibfnamefont {A.~K.}\ \bibnamefont {Robinson}}, \bibinfo {author}
  {\bibfnamefont {D.~S.}\ \bibnamefont {La~Mantia}},\ and\ \bibinfo {author}
  {\bibfnamefont {E.~B.}\ \bibnamefont {Norrgard}},\ }\bibfield  {title}
  {\bibinfo {title} {Electromagnetically induced transparency based
  rydberg-atom sensor for traceable voltage measurements},\ }\href
  {https://doi.org/10.1116/5.0097746} {\bibfield  {journal} {\bibinfo
  {journal} {AVS Quantum Science}\ }\textbf {\bibinfo {volume} {4}},\ \bibinfo
  {pages} {034401} (\bibinfo {year} {2022})}\BibitemShut {NoStop}%
\bibitem [{\citenamefont {Li}\ \emph {et~al.}(2023)\citenamefont {Li},
  \citenamefont {Jiao}, \citenamefont {Hu}, \citenamefont {Li}, \citenamefont
  {Shi}, \citenamefont {Zhao},\ and\ \citenamefont {Jia}}]{li2023}%
  \BibitemOpen
  \bibfield  {author} {\bibinfo {author} {\bibfnamefont {L.}~\bibnamefont
  {Li}}, \bibinfo {author} {\bibfnamefont {Y.}~\bibnamefont {Jiao}}, \bibinfo
  {author} {\bibfnamefont {J.}~\bibnamefont {Hu}}, \bibinfo {author}
  {\bibfnamefont {H.}~\bibnamefont {Li}}, \bibinfo {author} {\bibfnamefont
  {M.}~\bibnamefont {Shi}}, \bibinfo {author} {\bibfnamefont {J.}~\bibnamefont
  {Zhao}},\ and\ \bibinfo {author} {\bibfnamefont {S.}~\bibnamefont {Jia}},\
  }\bibfield  {title} {\bibinfo {title} {Super low-frequency electric field
  measurement based on rydberg atoms},\ }\href
  {https://doi.org/10.1364/OE.499244} {\bibfield  {journal} {\bibinfo
  {journal} {Opt. Express}\ }\textbf {\bibinfo {volume} {31}},\ \bibinfo
  {pages} {29228} (\bibinfo {year} {2023})}\BibitemShut {NoStop}%
\bibitem [{\citenamefont {Lei}\ and\ \citenamefont {Shi}(2024)}]{lei24}%
  \BibitemOpen
  \bibfield  {author} {\bibinfo {author} {\bibfnamefont {M.}~\bibnamefont
  {Lei}}\ and\ \bibinfo {author} {\bibfnamefont {M.}~\bibnamefont {Shi}},\
  }\bibfield  {title} {\bibinfo {title} {High sensitivity measurement of ulf,
  vlf, and lf fields with a rydberg-atom sensor},\ }\href
  {https://doi.org/10.1364/OL.539090} {\bibfield  {journal} {\bibinfo
  {journal} {Opt. Lett.}\ }\textbf {\bibinfo {volume} {49}},\ \bibinfo {pages}
  {5547} (\bibinfo {year} {2024})}\BibitemShut {NoStop}%
\bibitem [{\citenamefont {Arumugam}(2025)}]{arumugam2025}%
  \BibitemOpen
  \bibfield  {author} {\bibinfo {author} {\bibfnamefont {D.}~\bibnamefont
  {Arumugam}},\ }\bibfield  {title} {\bibinfo {title} {Stark modulated rydberg
  dissipative time crystals at room temperature applied to sub-khz electric
  field sensing},\ }\bibfield  {journal} {\bibinfo  {journal} {Scientific
  Reports}\ }\textbf {\bibinfo {volume} {15}},\ \href
  {https://doi.org/10.1038/s41598-025-19859-x} {10.1038/s41598-025-19859-x}
  (\bibinfo {year} {2025})\BibitemShut {NoStop}%
\bibitem [{\citenamefont {Han}\ \emph {et~al.}(2025)\citenamefont {Han},
  \citenamefont {He}, \citenamefont {Weng}, \citenamefont {Xu}, \citenamefont
  {Zhao},\ and\ \citenamefont {Wang}}]{han2025}%
  \BibitemOpen
  \bibfield  {author} {\bibinfo {author} {\bibfnamefont {Y.}~\bibnamefont
  {Han}}, \bibinfo {author} {\bibfnamefont {C.}~\bibnamefont {He}}, \bibinfo
  {author} {\bibfnamefont {Z.}~\bibnamefont {Weng}}, \bibinfo {author}
  {\bibfnamefont {P.}~\bibnamefont {Xu}}, \bibinfo {author} {\bibfnamefont
  {Y.}~\bibnamefont {Zhao}},\ and\ \bibinfo {author} {\bibfnamefont
  {T.}~\bibnamefont {Wang}},\ }\bibfield  {title} {\bibinfo {title} {Dc and
  power-frequency electric field measurement with rydberg-atom
  interferometry},\ }\href {https://doi.org/10.1063/5.0272159} {\bibfield
  {journal} {\bibinfo  {journal} {Applied Physics Letters}\ }\textbf {\bibinfo
  {volume} {127}},\ \bibinfo {pages} {024002} (\bibinfo {year}
  {2025})}\BibitemShut {NoStop}%
\bibitem [{\citenamefont {Xiao}\ \emph {et~al.}(2025)\citenamefont {Xiao},
  \citenamefont {Wei}, \citenamefont {Yan},\ and\ \citenamefont
  {Zhang}}]{xiao2025}%
  \BibitemOpen
  \bibfield  {author} {\bibinfo {author} {\bibfnamefont {D.}~\bibnamefont
  {Xiao}}, \bibinfo {author} {\bibfnamefont {X.}~\bibnamefont {Wei}}, \bibinfo
  {author} {\bibfnamefont {S.}~\bibnamefont {Yan}},\ and\ \bibinfo {author}
  {\bibfnamefont {H.}~\bibnamefont {Zhang}},\ }\bibfield  {title} {\bibinfo
  {title} {Electric field measurement using a three-photon double-dark-state
  model in rydberg atoms},\ }\href {https://doi.org/10.1063/5.0265975}
  {\bibfield  {journal} {\bibinfo  {journal} {AIP Advances}\ }\textbf {\bibinfo
  {volume} {15}},\ \bibinfo {pages} {075108} (\bibinfo {year}
  {2025})}\BibitemShut {NoStop}%
\bibitem [{\citenamefont {Fletcher}(2017)}]{flectcher2017}%
  \BibitemOpen
  \bibfield  {author} {\bibinfo {author} {\bibfnamefont {A.~T.}\ \bibnamefont
  {Fletcher}},\ }\href@noop {} {\bibinfo {title} {A study of alkali-resistant
  materials for use in atomic physics based systems}} (\bibinfo {year}
  {2017})\BibitemShut {NoStop}%
\bibitem [{\citenamefont {Jau}(2021{\natexlab{a}})}]{Jau2021r}%
  \BibitemOpen
  \bibfield  {author} {\bibinfo {author} {\bibfnamefont {Y.-Y.}\ \bibnamefont
  {Jau}},\ }\href@noop {} {\emph {\bibinfo {title} {Advanced Electric-Field
  Sensor and Imaging Technologies}}},\ \bibinfo {type} {Tech. Rep.}\ \bibinfo
  {number} {1406005}\ (\bibinfo  {institution} {Sandia National Labs},\
  \bibinfo {year} {December 2021})\BibitemShut {NoStop}%
\bibitem [{\citenamefont {Karaulanov}\ \emph {et~al.}(2009)\citenamefont
  {Karaulanov}, \citenamefont {Graf}, \citenamefont {English}, \citenamefont
  {Rochester}, \citenamefont {Rosen}, \citenamefont {Tsigutkin}, \citenamefont
  {Budker}, \citenamefont {Alexandrov}, \citenamefont {Balabas}, \citenamefont
  {Kimball}, \citenamefont {Narducci}, \citenamefont {Pustelny},\ and\
  \citenamefont {Yashchuk}}]{Karaulanov2009}%
  \BibitemOpen
  \bibfield  {author} {\bibinfo {author} {\bibfnamefont {T.}~\bibnamefont
  {Karaulanov}}, \bibinfo {author} {\bibfnamefont {M.~T.}\ \bibnamefont
  {Graf}}, \bibinfo {author} {\bibfnamefont {D.}~\bibnamefont {English}},
  \bibinfo {author} {\bibfnamefont {S.~M.}\ \bibnamefont {Rochester}}, \bibinfo
  {author} {\bibfnamefont {Y.~J.}\ \bibnamefont {Rosen}}, \bibinfo {author}
  {\bibfnamefont {K.}~\bibnamefont {Tsigutkin}}, \bibinfo {author}
  {\bibfnamefont {D.}~\bibnamefont {Budker}}, \bibinfo {author} {\bibfnamefont
  {E.~B.}\ \bibnamefont {Alexandrov}}, \bibinfo {author} {\bibfnamefont
  {M.~V.}\ \bibnamefont {Balabas}}, \bibinfo {author} {\bibfnamefont
  {D.~F.~J.}\ \bibnamefont {Kimball}}, \bibinfo {author} {\bibfnamefont
  {F.~A.}\ \bibnamefont {Narducci}}, \bibinfo {author} {\bibfnamefont
  {S.}~\bibnamefont {Pustelny}},\ and\ \bibinfo {author} {\bibfnamefont
  {V.~V.}\ \bibnamefont {Yashchuk}},\ }\bibfield  {title} {\bibinfo {title}
  {Controlling atomic vapor density in paraffin-coated cells using
  light-induced atomic desorption},\ }\href
  {https://doi.org/10.1103/PhysRevA.79.012902} {\bibfield  {journal} {\bibinfo
  {journal} {Phys. Rev. A}\ }\textbf {\bibinfo {volume} {79}},\ \bibinfo
  {pages} {012902} (\bibinfo {year} {2009})}\BibitemShut {NoStop}%
\bibitem [{\citenamefont {Ma}\ \emph {et~al.}(2025)\citenamefont {Ma},
  \citenamefont {Xiao}, \citenamefont {Zhang}, \citenamefont {Wang},\ and\
  \citenamefont {Wei}}]{ma2025}%
  \BibitemOpen
  \bibfield  {author} {\bibinfo {author} {\bibfnamefont {K.}~\bibnamefont
  {Ma}}, \bibinfo {author} {\bibfnamefont {D.}~\bibnamefont {Xiao}}, \bibinfo
  {author} {\bibfnamefont {H.}~\bibnamefont {Zhang}}, \bibinfo {author}
  {\bibfnamefont {X.}~\bibnamefont {Wang}},\ and\ \bibinfo {author}
  {\bibfnamefont {X.}~\bibnamefont {Wei}},\ }\bibfield  {title} {\bibinfo
  {title} {Study on electric field shielding in sio2 and caf2 vapor cell for
  rydberg atom electric field sensors},\ }\href
  {https://doi.org/10.1088/1361-6463/adb04c} {\bibfield  {journal} {\bibinfo
  {journal} {Journal of Physics D: Applied Physics}\ }\textbf {\bibinfo
  {volume} {58}},\ \bibinfo {pages} {135113} (\bibinfo {year}
  {2025})}\BibitemShut {NoStop}%
\bibitem [{\citenamefont {Jau}\ and\ \citenamefont
  {Christensen}(2022)}]{Jau2022r}%
  \BibitemOpen
  \bibfield  {author} {\bibinfo {author} {\bibfnamefont {Y.-Y.}\ \bibnamefont
  {Jau}}\ and\ \bibinfo {author} {\bibfnamefont {J.}~\bibnamefont
  {Christensen}},\ }\href@noop {} {\emph {\bibinfo {title} {Better quasi-DC
  atomic electrometer and new magnetoresistance}}},\ \bibinfo {type} {Tech.
  Rep.}\ \bibinfo {number} {SD16136}\ (\bibinfo  {institution} {Sandia National
  Labs},\ \bibinfo {year} {April 2022})\BibitemShut {NoStop}%
\bibitem [{\citenamefont {Ramirez}(1997)}]{ramirez1997}%
  \BibitemOpen
  \bibfield  {author} {\bibinfo {author} {\bibfnamefont {A.~P.}\ \bibnamefont
  {Ramirez}},\ }\bibfield  {title} {\bibinfo {title} {Colossal
  magnetoresistance},\ }\href {https://doi.org/10.1088/0953-8984/9/39/005}
  {\bibfield  {journal} {\bibinfo  {journal} {Journal of Physics: Condensed
  Matter}\ }\textbf {\bibinfo {volume} {9}},\ \bibinfo {pages} {8171} (\bibinfo
  {year} {1997})}\BibitemShut {NoStop}%
\bibitem [{\citenamefont {Ennen}\ \emph {et~al.}(2016)\citenamefont {Ennen},
  \citenamefont {Kappe}, \citenamefont {Rempel}, \citenamefont {Glenske},\ and\
  \citenamefont {Hütten}}]{ennen2016}%
  \BibitemOpen
  \bibfield  {author} {\bibinfo {author} {\bibfnamefont {I.}~\bibnamefont
  {Ennen}}, \bibinfo {author} {\bibfnamefont {D.}~\bibnamefont {Kappe}},
  \bibinfo {author} {\bibfnamefont {T.}~\bibnamefont {Rempel}}, \bibinfo
  {author} {\bibfnamefont {C.}~\bibnamefont {Glenske}},\ and\ \bibinfo {author}
  {\bibfnamefont {A.}~\bibnamefont {Hütten}},\ }\bibfield  {title} {\bibinfo
  {title} {Giant magnetoresistance: Basic concepts, microstructure, magnetic
  interactions and applications},\ }\bibfield  {journal} {\bibinfo  {journal}
  {Sensors}\ }\textbf {\bibinfo {volume} {16}},\ \href
  {https://doi.org/10.3390/s16060904} {10.3390/s16060904} (\bibinfo {year}
  {2016})\BibitemShut {NoStop}%
\bibitem [{\citenamefont {Niu}\ and\ \citenamefont {Zhu}(2021)}]{niu2022}%
  \BibitemOpen
  \bibfield  {author} {\bibinfo {author} {\bibfnamefont {R.}~\bibnamefont
  {Niu}}\ and\ \bibinfo {author} {\bibfnamefont {W.~K.}\ \bibnamefont {Zhu}},\
  }\bibfield  {title} {\bibinfo {title} {Materials and possible mechanisms of
  extremely large magnetoresistance: a review},\ }\href
  {https://doi.org/10.1088/1361-648X/ac3b24} {\bibfield  {journal} {\bibinfo
  {journal} {Journal of Physics: Condensed Matter}\ }\textbf {\bibinfo {volume}
  {34}},\ \bibinfo {pages} {113001} (\bibinfo {year} {2021})}\BibitemShut
  {NoStop}%
\bibitem [{\citenamefont {Ritzinger}\ and\ \citenamefont
  {Výborný}(2023)}]{ritzinger2023}%
  \BibitemOpen
  \bibfield  {author} {\bibinfo {author} {\bibfnamefont {P.}~\bibnamefont
  {Ritzinger}}\ and\ \bibinfo {author} {\bibfnamefont {K.}~\bibnamefont
  {Výborný}},\ }\bibfield  {title} {\bibinfo {title} {Anisotropic
  magnetoresistance: materials, models and applications},\ }\href
  {https://doi.org/10.1098/rsos.230564} {\bibfield  {journal} {\bibinfo
  {journal} {Royal Society Open Science}\ }\textbf {\bibinfo {volume} {10}},\
  \bibinfo {pages} {230564} (\bibinfo {year} {2023})}\BibitemShut {NoStop}%
\bibitem [{Wor()}]{WorkFunctionNote}%
  \BibitemOpen
  \href@noop {} {}\bibinfo {note} {In the 2020 experiment on LED-induced
  E-field biasing inside the vapor cell at SNL, we tested LED light with
  wavelengths ranging from 370 nm to 980 nm. The internal E-field generation
  from the optically excited charge patches became noticeable only when the
  wavelength was shorter than approximately 600 nm.}\BibitemShut {Stop}%
\bibitem [{\citenamefont {Ryabtsev}\ \emph {et~al.}(2011)\citenamefont
  {Ryabtsev}, \citenamefont {Beterov}, \citenamefont {Tretyakov}, \citenamefont
  {Entin},\ and\ \citenamefont {Yakshina}}]{Ryabtsev2011}%
  \BibitemOpen
  \bibfield  {author} {\bibinfo {author} {\bibfnamefont {I.~I.}\ \bibnamefont
  {Ryabtsev}}, \bibinfo {author} {\bibfnamefont {I.~I.}\ \bibnamefont
  {Beterov}}, \bibinfo {author} {\bibfnamefont {D.~B.}\ \bibnamefont
  {Tretyakov}}, \bibinfo {author} {\bibfnamefont {V.~M.}\ \bibnamefont
  {Entin}},\ and\ \bibinfo {author} {\bibfnamefont {E.~A.}\ \bibnamefont
  {Yakshina}},\ }\bibfield  {title} {\bibinfo {title} {Doppler- and recoil-free
  laser excitation of rydberg states via three-photon transitions},\ }\href
  {https://doi.org/10.1103/PhysRevA.84.053409} {\bibfield  {journal} {\bibinfo
  {journal} {Phys. Rev. A}\ }\textbf {\bibinfo {volume} {84}},\ \bibinfo
  {pages} {053409} (\bibinfo {year} {2011})}\BibitemShut {NoStop}%
\bibitem [{\citenamefont {Ripka}\ \emph {et~al.}(2021)\citenamefont {Ripka},
  \citenamefont {Amarloo}, \citenamefont {Erskine}, \citenamefont {Liu},
  \citenamefont {Ramirez-Serrano}, \citenamefont {Keaveney}, \citenamefont
  {Gillet}, \citenamefont {Kubler},\ and\ \citenamefont {Shaffer}}]{Ripka2021}%
  \BibitemOpen
  \bibfield  {author} {\bibinfo {author} {\bibfnamefont {F.}~\bibnamefont
  {Ripka}}, \bibinfo {author} {\bibfnamefont {H.}~\bibnamefont {Amarloo}},
  \bibinfo {author} {\bibfnamefont {J.}~\bibnamefont {Erskine}}, \bibinfo
  {author} {\bibfnamefont {C.}~\bibnamefont {Liu}}, \bibinfo {author}
  {\bibfnamefont {J.}~\bibnamefont {Ramirez-Serrano}}, \bibinfo {author}
  {\bibfnamefont {J.}~\bibnamefont {Keaveney}}, \bibinfo {author}
  {\bibfnamefont {G.}~\bibnamefont {Gillet}}, \bibinfo {author} {\bibfnamefont
  {H.}~\bibnamefont {Kubler}},\ and\ \bibinfo {author} {\bibfnamefont {J.~P.}\
  \bibnamefont {Shaffer}},\ }\bibfield  {title} {\bibinfo {title}
  {Application-driven problems in rydberg atom electrometry},\ }\bibinfo
  {organization} {SPIE}\ (\bibinfo  {publisher} {Proc. of SPIE},\ \bibinfo
  {year} {2021})\ p.\ \bibinfo {pages} {117002Y}\BibitemShut {NoStop}%
\bibitem [{\citenamefont {Jau}(2021{\natexlab{b}})}]{Jau2021d}%
  \BibitemOpen
  \bibfield  {author} {\bibinfo {author} {\bibfnamefont {Y.-Y.}\ \bibnamefont
  {Jau}},\ }\href@noop {} {\emph {\bibinfo {title} {Modeling for atomic sensor
  improvements}}},\ \bibinfo {type} {Tech. Rep.}\ \bibinfo {number} {1305650}\
  (\bibinfo  {institution} {Sandia National Labs},\ \bibinfo {year} {April
  2021})\BibitemShut {NoStop}%
\bibitem [{\citenamefont {Happer}\ \emph {et~al.}(2010)\citenamefont {Happer},
  \citenamefont {Jau},\ and\ \citenamefont {Walker}}]{Happer2010}%
  \BibitemOpen
  \bibfield  {author} {\bibinfo {author} {\bibfnamefont {W.}~\bibnamefont
  {Happer}}, \bibinfo {author} {\bibfnamefont {Y.-Y.}\ \bibnamefont {Jau}},\
  and\ \bibinfo {author} {\bibfnamefont {T.}~\bibnamefont {Walker}},\
  }\href@noop {} {\emph {\bibinfo {title} {Optically Pumped Atoms}}}\ (\bibinfo
   {publisher} {Wiley-VCH},\ \bibinfo {year} {2010})\BibitemShut {NoStop}%
\bibitem [{\citenamefont {Jau}\ \emph {et~al.}(2016)\citenamefont {Jau},
  \citenamefont {Hankin}, \citenamefont {Keating}, \citenamefont {Deutsch},\
  and\ \citenamefont {Biedermann}}]{Jau2016}%
  \BibitemOpen
  \bibfield  {author} {\bibinfo {author} {\bibfnamefont {Y.-Y.}\ \bibnamefont
  {Jau}}, \bibinfo {author} {\bibfnamefont {A.~M.}\ \bibnamefont {Hankin}},
  \bibinfo {author} {\bibfnamefont {T.}~\bibnamefont {Keating}}, \bibinfo
  {author} {\bibfnamefont {I.~H.}\ \bibnamefont {Deutsch}},\ and\ \bibinfo
  {author} {\bibfnamefont {G.~W.}\ \bibnamefont {Biedermann}},\ }\bibfield
  {title} {\bibinfo {title} {Entangling atomic spins with a {R}ydberg-dressed
  spin-flip blockade},\ }\href@noop {} {\bibfield  {journal} {\bibinfo
  {journal} {Nat. Phys.}\ }\textbf {\bibinfo {volume} {12}},\ \bibinfo {pages}
  {71} (\bibinfo {year} {2016})}\BibitemShut {NoStop}%
\bibitem [{\citenamefont {Jau}(2024)}]{Jau2024r}%
  \BibitemOpen
  \bibfield  {author} {\bibinfo {author} {\bibfnamefont {Y.-Y.}\ \bibnamefont
  {Jau}},\ }\href@noop {} {\emph {\bibinfo {title} {Field-deployable quasi-DC
  atomic E-field sensor system}}},\ \bibinfo {type} {Tech. Rep.}\ \bibinfo
  {number} {1759873}\ (\bibinfo  {institution} {Sandia National Labs},\
  \bibinfo {year} {October 2024})\BibitemShut {NoStop}%
\bibitem [{Vid()}]{Video}%
  \BibitemOpen
  \href@noop {} {}\bibinfo {note} {See supplemental materials for a video
  illustration.}\BibitemShut {Stop}%
\bibitem [{\citenamefont {Jau}()}]{ElectronicModel}%
  \BibitemOpen
  \bibfield  {author} {\bibinfo {author} {\bibfnamefont {Y.-Y.}\ \bibnamefont
  {Jau}},\ }\href@noop {} {}\bibinfo {note} {Private manuscript, ``Notes on the
  thermal E-field noise inside the vapor cell, electronic EM-field detection,
  and fundamental limits,'' Section of Electronic E-field Detection
  (2021)}\BibitemShut {NoStop}%
\bibitem [{\citenamefont {Krupka}\ \emph {et~al.}(2001)\citenamefont {Krupka},
  \citenamefont {Matthews}, \citenamefont {Say},\ and\ \citenamefont
  {Hibbs}}]{Krupka2001}%
  \BibitemOpen
  \bibfield  {author} {\bibinfo {author} {\bibfnamefont {M.~A.}\ \bibnamefont
  {Krupka}}, \bibinfo {author} {\bibfnamefont {R.}~\bibnamefont {Matthews}},
  \bibinfo {author} {\bibfnamefont {C.}~\bibnamefont {Say}},\ and\ \bibinfo
  {author} {\bibfnamefont {A.}~\bibnamefont {Hibbs}},\ }\href@noop {} {\emph
  {\bibinfo {title} {Development and Test of Free Space Electric Field Sensors
  with Microvolt Sensitivity}}},\ \bibinfo {type} {Technical Report}\ (\bibinfo
   {institution} {Berkeley Space Sciences Laboratory},\ \bibinfo {year}
  {2001})\BibitemShut {NoStop}%
\bibitem [{\citenamefont {Gurses}\ \emph {et~al.}(2021)\citenamefont {Gurses},
  \citenamefont {Whitmore},\ and\ \citenamefont {Cohen}}]{Gurses2021}%
  \BibitemOpen
  \bibfield  {author} {\bibinfo {author} {\bibfnamefont {B.~V.}\ \bibnamefont
  {Gurses}}, \bibinfo {author} {\bibfnamefont {K.~T.}\ \bibnamefont
  {Whitmore}},\ and\ \bibinfo {author} {\bibfnamefont {M.~B.}\ \bibnamefont
  {Cohen}},\ }\bibfield  {title} {\bibinfo {title} {Ultra-sensitive broadband
  “awesome” electric field receiver for nanovolt low-frequency signals},\
  }\href@noop {} {\bibfield  {journal} {\bibinfo  {journal} {Rev. Sci.
  Instrum.}\ }\textbf {\bibinfo {volume} {92}},\ \bibinfo {pages} {024704}
  (\bibinfo {year} {2021})}\BibitemShut {NoStop}%
\bibitem [{ARL()}]{ARL2026}%
  \BibitemOpen
  \href@noop {} {}\bibinfo {note} {Private communication with Dr. Paul Kunz and
  his associated research teams.}\BibitemShut {Stop}%
\bibitem [{\citenamefont {Kayim}\ \emph {et~al.}(2026)\citenamefont {Kayim},
  \citenamefont {Viray}, \citenamefont {Mantia}, \citenamefont {Richardson},
  \citenamefont {Dee}, \citenamefont {Westafer}, \citenamefont {Sawyer},\ and\
  \citenamefont {Wyllie}}]{Kayim2026}%
  \BibitemOpen
  \bibfield  {author} {\bibinfo {author} {\bibfnamefont {B.}~\bibnamefont
  {Kayim}}, \bibinfo {author} {\bibfnamefont {M.~A.}\ \bibnamefont {Viray}},
  \bibinfo {author} {\bibfnamefont {D.~S.~L.}\ \bibnamefont {Mantia}}, \bibinfo
  {author} {\bibfnamefont {D.}~\bibnamefont {Richardson}}, \bibinfo {author}
  {\bibfnamefont {J.}~\bibnamefont {Dee}}, \bibinfo {author} {\bibfnamefont
  {R.~S.}\ \bibnamefont {Westafer}}, \bibinfo {author} {\bibfnamefont {B.~C.}\
  \bibnamefont {Sawyer}},\ and\ \bibinfo {author} {\bibfnamefont
  {R.}~\bibnamefont {Wyllie}},\ }\bibfield  {title} {\bibinfo {title}
  {Calibration of electric fields in low-frequency off-resonant rydberg
  receivers},\ }\href@noop {} {\bibfield  {journal} {\bibinfo  {journal}
  {arXiv}\ ,\ \bibinfo {pages} {2603.10898}} (\bibinfo {year}
  {2026})}\BibitemShut {NoStop}%
\bibitem [{\citenamefont {Chandra}\ \emph {et~al.}(2026)\citenamefont
  {Chandra}, \citenamefont {Paensin},\ and\ \citenamefont
  {Dumke}}]{Chandra2026}%
  \BibitemOpen
  \bibfield  {author} {\bibinfo {author} {\bibfnamefont {A.}~\bibnamefont
  {Chandra}}, \bibinfo {author} {\bibfnamefont {N.}~\bibnamefont {Paensin}},\
  and\ \bibinfo {author} {\bibfnamefont {R.}~\bibnamefont {Dumke}},\ }\bibfield
   {title} {\bibinfo {title} {Electrometry of extremely-low frequencies from
  khz to sub-hz with a rydberg-atom sensor},\ }\href@noop {} {\bibfield
  {journal} {\bibinfo  {journal} {arXiv}\ ,\ \bibinfo {pages} {2603.13827}}
  (\bibinfo {year} {2026})}\BibitemShut {NoStop}%
\end{thebibliography}%

\end{document}